\documentclass[ip,twocolumn]{jpsj3} 
\usepackage{txfonts}
\usepackage[usenames]{color}
\usepackage{bm}
\usepackage{braket}

\title{A New Era of Excitonic Insulators}
\author{Tatsuya Kaneko$^1$ and Yukinori Ohta$^2$}
\inst{$^1$Department of Physics, Osaka University, Toyonaka, Osaka 560-0043, Japan \\
$^2$Department of Physics, Chiba University, Chiba 263-8522, Japan} 
\abst{The fundamental idea of the excitonic insulator (EI) driven by electron--hole correlations in narrow-gap semiconductors or semimetals was originally proposed in the 1960s, and only theoretical studies had been advanced for a long time. 
However, the rise of new candidate materials and recent developments in measurement techniques have enabled us to discuss the possibilities of EI states in real materials experimentally.  
In this article, we review recent progress in the research of EIs. 
We start with an introduction to the theoretical background of the EI and the mechanism of the order formation including its relation to the physics of the Bardeen--Cooper--Schrieffer (BCS) -- Bose--Einstein condensation (BEC) crossover. 
We also review the EI states studied in the context of strongly correlated electron systems. 
Then, we introduce the candidate materials for the EI and the issues raised by recent experiments. 
For example, the phase transitions in several candidate materials are accompanied by lattice distortions, and the contributions from electron--lattice coupling hinder the identification of the excitonic entity. 
We also review the profiles of the collective modes to discuss the dynamical signatures of the EI.}

\begin{document}

\maketitle


\section{Introduction}

Electrons in solids sometimes show phase transitions into an ordered state at low temperatures, which create intriguing material properties~\cite{sachdev2011,khomskii2014,keimer2015}.  
For example, the ordered states of spins give rise to magnetism in materials, and phase transitions into superconducting states formed by electron--electron pairs markedly change transport properties. 
Phase transition caused by ordered electrons is a central issue in condensed matter physics, and its applications have afforded many functions to materials. 

In the 1960s after the establishment of the Bardeen--Cooper--Schrieffer (BCS) theory of superconductivity~\cite{bardeen1957}, the idea of the excitonic insulator (EI) was theoretically proposed as an ordered state of electron--hole pairs~\cite{mott1961,knox1963,keldysh1965,descloizeaux1965,kozlov1965,jerome1967,zittartz1967,kohn1967,halperin1968,halperin1968_2}. 
The EI state is driven by interband Coulomb (i.e., excitonic) interactions in narrow-gap semiconductors and band-overlapped semimetals, and its phase transition reconfigures the band structures at low temperatures. 
Although the components of pairs and the pairing glue are quite different from the conventional superconducting order, the mathematical structures including the gap equation of the excitonic order are analogous to those in the theory of superconductivity.
Therefore, the formation of the excitonic order is often discussed in association with pair condensations~\cite{kohn1970,moskalenko2000,littlewood2004}.  
Given such background, theoretical studies of the EI have been conducted since the 1960s.   
On the other hand, not much experimental research on the EI was conducted for a long time in part because no promising signature of an excitonic order was found in real materials.  

However, developments in measurement techniques and the rise of new candidate materials have enabled us to discuss EI states experimentally in the last few decades. 
The candidate material TiSe$_2$~\cite{traum1978} has been investigated in detail using, for example, angle-resolved photoemission spectroscopy and momentum-resolved electron energy loss spectroscopy~\cite{cercellier2007,kogar2017}.  
The quasi-one-dimensional material Ta$_2$NiSe$_5$ has also been discussed as a candidate~\cite{wakisaka2009,seki2014,lu2017}, where various optical experiments, including pump-probe spectroscopy, have been conducted~\cite{werdehausen2018,kim2021}. 
Regarding spin-involved materials, the cobalt oxides near the spin-state crossover regime have been studied as hosts of the spin-triplet-type EI~\cite{kunes2014Co,ikeda2023}. 
Recently, a possible EI state has been suggested in monolayer WTe$_2$, which was originally studied as a two-dimensional (2D) topological insulator~\cite{jia2022,sun2022}. 
The recent developments in the studies of the EIs have also been encouraged by modern theoretical techniques.  
The development of band structure calculations based on the density functional theory (DFT) has enabled us to assess realistic electronic structures of materials. 
Moreover, solvers for quantum many-body problems, such as dynamical mean-field theory (DMFT)~\cite{georges1996} and the density-matrix renormalization group (DMRG) method~\cite{white1992}, have made it possible to evaluate strongly correlated states precisely beyond the simple single-particle approximations.  
As a result of these developments, the research of the EI has proceeded to the stage where we can combine theoretical predictions with state-of-the-art experiments.  
Although there is still much debate, the fact that we can now combine theory and experiment to discuss the possibilities of EI states is an important advancement in the last few decades. 

In the debate about the experimental realization of the EI state, identifying the contributions from the excitonic correlations originating from interband interactions often becomes a critical issue.  
In particular, the phase transitions in TiSe$_2$ and Ta$_2$NiSe$_5$ are accompanied by lattice distortions, which hinder the detection of the excitonic entities.    
Because both excitonic interactions and electron--phonon couplings can contribute to the bandgap opening, it is often difficult to resolve these two contributions in simple single-particle bands.  
In these systems, understanding the collective dynamics of EIs coupled to lattice systems becomes important in distinguishing between the lattice and excitonic contributions because the phonons and the collective electronic motions usually possess different dynamical properties. 
Recent pump-probe experiments have enabled us to obtain the nonequilibrium dynamics of electron--phonon coupled systems, whereas its accurate interpretation is desired.    
As seen in this example, although various candidate materials have been proposed recently, active discussions are ongoing regarding each material.  

In this article, we review the recent progress in the research of the EI. 
In Sect.~\ref{sec:intro_EI}, we introduce the theoretical background.  
We briefly review the mechanism of the gap opening in the excitonic order and provide insights related to the crossover between the weak-coupling BCS-like regime and the strong-coupling Bose--Einstein condensation (BEC)-like regime of the order formation. 
In Sect.~\ref{sec:EI_SCES}, we review the EI states from the perspective of strongly correlated electron systems, where we introduce the theoretical studies using the spinless Falicov--Kimball model and the two-orbital Hubbard model.  
In Sect.~\ref{sec:candidates}, we introduce the candidate materials and various experimental studies on them. 
In particular, we feature the candidate materials TiSe$_2$ and Ta$_2$NiSe$_5$.
We also pick up the cobalt oxides as candidates for spin-triplet-type excitonic orders. 
In addition, we briefly introduce other candidate materials proposed to date. 
In Sect.~\ref{sec:CM}, we review the character of the collective modes.   
We consider the collective modes in the pure excitonic order and then discuss the effects of electron--lattice coupling. 
In Sect.~\ref{sec:summary}, we summarize our review.


\section{Excitonic Order} \label{sec:intro_EI}

First, we provide an overview of the theoretical background of the EI. 
Using a simple two-band model, we introduce the order parameter of the excitonic order and the resulting gap equation. 
Then, we review a typical phase diagram, where the physics of the EI in the range from narrow-gap semiconductors to band-overlapped semimetals is discussed in association with the BCS--BEC crossover. 
Since electrons in actual materials have spin degrees of freedom, we also introduce the excitonic orders considering spin.

\subsection{Order parameter and gap equation} \label{sec:EO}

\begin{figure}[t]
\begin{center}  
\includegraphics[width=\columnwidth]{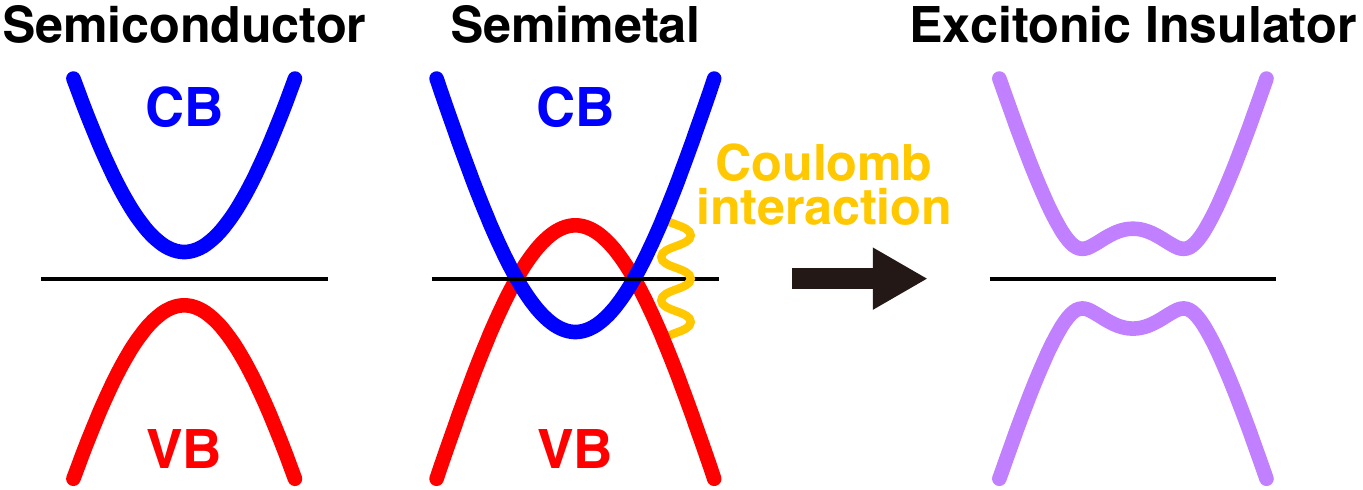}          
\caption{(Color online) Semiconductor, semimetal, and excitonic insulator.}
\label{fig1}
\end{center}
\end{figure}

Employing a simple two-band correlated model, we introduce the essence of the theory.  
Here, we consider the Hamiltonian $\hat{\mathcal{H}}=\hat{\mathcal{H}}_0+\hat{\mathcal{H}}_V$ consisting of the noninteracting term 
\begin{align}
\hat{\mathcal{H}}_0 
&= \sum_{i,j}  t^{(a)}_{ij} \hat{a}^{\dag}_{i} \hat{a}_{j} + \sum_{j}  \epsilon^{(a)} \hat{a}^{\dag}_{j} \hat{a}_{j}
+ \sum_{i,j}  t^{(b)}_{ij} \hat{b}^{\dag}_{i} \hat{b}_{j} + \sum_{j}  \epsilon^{(a)} \hat{b}^{\dag}_{j} \hat{b}_{j}
\notag \\
&= \sum_{\bm{k}} \varepsilon_a(\bm{k}) \hat{a}^{\dag}_{\bm{k}} \hat{a}_{\bm{k}}
+  \sum_{\bm{k}} \varepsilon_b(\bm{k}) \hat{b}^{\dag}_{\bm{k}} \hat{b}_{\bm{k}}
\label{eq:ham_twoband_0}
\end{align}
and the local interaction term 
\begin{equation}
\hat{\mathcal{H}}_V = V \sum_{j}  \hat{a}^{\dag}_{j} \hat{a}_j \hat{b}^{\dag}_{j} \hat{b}_{j}. 
\label{eq:ham_twoband_V}
\end{equation}
$\hat{a}_j$ ($\hat{a}^{\dag}_j$) and $\hat{b}_j$ ($\hat{b}^{\dag}_j$) are the annihilation (creation) operators of fermions at site $j$ on the $a$ and $b$ orbitals, respectively. 
$\hat{a}_{\bm{k}}$ ($\hat{a}^{\dag}_{\bm{k}}$) and $\hat{b}_{\bm{k}}$ ($\hat{b}^{\dag}_{\bm{k}}$) are the annihilation (creation) operators in their reciprocal ($\bm{k}$) space. 
Here, we consider the spinless fermions for simplicity.   
We will introduce EI states with spin degrees of freedom in Sect.~\ref{sec:EI_spin}.  
$t^{(\gamma)}_{ij}$ is the hopping integral between the $i$ and $j$ sites on the $\gamma$ ($= a,b$) orbital, and $\epsilon^{(\gamma)}$ is the energy level of the $\gamma$ orbital.  
We set $\epsilon^{(a)} > \epsilon^{(b)}$, i.e., $\varepsilon_a(\bm{k})$ and $\varepsilon_b(\bm{k})$ correspond to the conduction band (CB) composed of orbital $a$ and the valence band (VB) composed of orbital $b$, respectively.  
We assume there is no hybridization between the two bands in the noninteracting system. 
[Note that hybridization can be generated by several causes, such as electron--lattice coupling (see Sect.~\ref{sec:CM}), but we neglect it for simplicity in this section.] 
For example, $\varepsilon_{\gamma}(\bm{k})=\epsilon^{(\gamma)} + 2t^{(\gamma)}(\cos k_x + \cos k_y )$ on the 2D square lattice with the nearest-neighbor hopping $t^{(\gamma)}$ (where the lattice constant is set to 1). 
Figure~\ref{fig1} (left) shows band structures of a semiconductor and a semimetal around the Fermi level. 
Here, we assume that the top of the VB and the bottom of the CB are located at the same $\bm{k}$ point (when $t^{(a)} t^{(b)}<0$). 
$V$ corresponds to the interband Coulomb interaction, which induces the excitonic correlation in this model. 
Although we only consider the local interaction for simplicity, we can form a similar gap equation even if the interaction is nonlocal (see below). 
In terms of a correlated lattice model, the spinless extended Falicov--Kimball model~\cite{batista2002} gives the same Hamiltonian (see Sect.~\ref{sec:EI_SCES_spinless}). 

The interaction $V$ between the VB and CB can modify the many-body state. 
To incorporate the effect of $V$ most simply, we apply the Hartree--Fock approximation to the interaction term $\hat{\mathcal{H}}_V$. 
This approximation gives the Hartree term, such as $V\braket{\hat{a}^{\dag}_j \hat{a}_j}\hat{b}^{\dag}_j \hat{b}_j$, which leads to the energy shift depending on the occupancy of each orbital, and the Fock term, such as $-V\braket{\hat{a}^{\dag}_j \hat{b}_j}\hat{b}^{\dag}_j \hat{a}_j$, which gives rise to hybridization.  
Here, the expectation value $\braket{\hat{a}^{\dag}_j \hat{b}_j}$ in the Fock term corresponds to the order parameter of the excitonic order. 
The operator $\hat{a}^{\dag}_j \hat{b}_j$ indicates the creation of an electron--hole pair, and the spontaneous generation of its expectation value $\braket{\hat{a}^{\dag}_j \hat{b}_j}$ can be regarded as the order formation (or condensation) of electron--hole pairs. 
For $\hat{\mathcal{H}_V}$ in Eq.~(\ref{eq:ham_twoband_V}), we may define the order parameter $\Delta$ as 
\begin{equation}
\Delta = -\frac{V}{N} \sum_{\bm{k}} \braket{ \hat{a}^{\dag}_{\bm{k}} \hat{b}_{\bm{k}} }, 
\label{eq:EIOP}
\end{equation}
where $N$ is the number of unit cells. 
In this definition, $\hat{a}^{\dag}_{\bm{k}}$ represents the creation of an electron on the CB, whereas $\hat{b}_{\bm{k}}$ represents the creation of a hole on the VB. 
This is the order parameter when the VB top and CB bottom are located at the same $\bm{k}$ point, as shown in Fig.~\ref{fig1} (left). 
Then, for the vector $\hat{\bm{\Psi}}_{\bm{k}}^{\dag} = (\hat{a}^{\dag}_{\bm{k}} \: \hat{b}^{\dag}_{\bm{k}})$, the approximated Hamiltonian (except for the constant term) is represented by $\hat{\mathcal{H}}_{\rm HF} = \sum_{\bm{k}} \hat{\bm{\Psi}}_{\bm{k}}^{\dag}  \mathcal{H}_{\bm{k}} \hat{\bm{\Psi}}_{\bm{k}}$ with the matrix 
\begin{equation}
\mathcal{H}_{\bm{k}}= \left(
\begin{array}{cc} 
  \bar{\varepsilon}_a(\bm{k}) & \Delta^* \\
  \Delta & \bar{\varepsilon}_b(\bm{k}) 
\end{array} 
\right), 
\label{eq:hammf_twoband_matrix}
\end{equation}
where $\bar{\varepsilon}_a(\bm{k}) $ and $\bar{\varepsilon}_b(\bm{k}) $ in the diagonal components are the energies including the Hartree shift. 
When the occupancy of the $\gamma$~$(=a,b)$ orbital is denoted as $n_{\gamma}$, $\bar{\varepsilon}_a(\bm{k}) = \varepsilon_a(\bm{k}) + V n_b$ and $\bar{\varepsilon}_b(\bm{k}) = \varepsilon_b(\bm{k}) + V n_a$. 
The order parameter $\Delta$ is in the off-diagonal component. 
If $\Delta = 0$, there is no hybridization and the two bands are independent. 
A nonzero order parameter $\Delta \ne 0$ leads to hybridization and gives the renewed eigenenergies
\begin{equation}
E_{\pm}(\bm{k}) = \frac{ \bar{\varepsilon}_a(\bm{k}) + \bar{\varepsilon}_b(\bm{k}) }{2}  \pm \sqrt{ \left( \frac{ \bar{\varepsilon}_a(\bm{k}) - \bar{\varepsilon}_b(\bm{k}) }{2} \right)^2  + |\Delta|^2}. 
\label{eq:EI_eigenenergy}
\end{equation}
These energies give the band structure shown in Fig.~\ref{fig1} (right), where band hybridization due to $\Delta \ne 0$ opens a gap between the VB and CB. 
Because of this gap opening, the simple two-band system becomes insulating, i.e., the EI state is realized. 

Similarly to the BCS theory of superconductivity, the presence of the order is judged from the self-consistent gap equation. 
By transforming the expectation value $\braket{\hat{a}^{\dag}_{\bm{k}} \hat{b}_{\bm{k}}}$ in Eq.~(\ref{eq:EIOP}) using the eigenvectors of Eq.~(\ref{eq:hammf_twoband_matrix}), we can derive the gap equation 
\begin{equation}
1 = -\frac{V}{N} \sum_{\bm{k}} \frac{1}{2\sqrt{\xi^2_{\bm{k}} + |\Delta|^2}} \left[ f(E_+(\bm{k})) - f(E_-(\bm{k}))\right].  
\label{eq:Gapeq}
\end{equation} 
$\xi_{\bm{k}} = [ \bar{\varepsilon}_a(\bm{k}) - \bar{\varepsilon}_b(\bm{k}) ] /2$ and $f(E)=1/[e^{\beta(E-\mu)}+1]$ is the Fermi distribution function (where $\beta$ is inverse temperature and  $\mu$ is chemical potential). 
If $\bar{\varepsilon}_a(\bm{k}) = - \bar{\varepsilon}_b(\bm{k})$, i.e., the system is particle-hole symmetric, Eq.~(\ref{eq:Gapeq}) becomes a similar form to the gap equation in the BCS theory. 

Although only the local interaction is considered in Eq.~(\ref{eq:ham_twoband_V}), the Coulomb interaction is generally nonlocal. 
When the nonlocal contributions are considered, the interband interaction has $\bm{k}$-dependence in the reciprocal space. 
Then, the order parameter is defined as 
\begin{equation}
\Delta(\bm{k}) = -\frac{1}{N} \sum_{\bm{k}'} V(\bm{k}-\bm{k'}) \braket{ \hat{a}^{\dag}_{\bm{k}'} \hat{b}_{\bm{k}'} }, 
\label{eq:EIOP_Vk}
\end{equation}
where $V(\bm{k}-\bm{k'})$ described in the momentum space is the interband Coulomb interaction including the nonlocal contributions.   
The order parameter $\Delta(\bm{k})$ has $\bm{k}$-dependence, and the gap equation is given by~\cite{kozlov1965,jerome1967,zittartz1967}
\begin{equation}
\Delta(\bm{k}) = -\frac{1}{N} \sum_{\bm{k}'} \frac{V(\bm{k}-\bm{k'})\Delta(\bm{k}')}{2\sqrt{\xi^2_{\bm{k}'} + |\Delta(\bm{k}')|^2}} \left[ f(E_+(\bm{k}')) - f(E_-(\bm{k}'))\right]. 
\label{eq:Gapeq_Vk}
\end{equation}
The solution of this gap equation has been provided in the early stage of studies, such as in Refs.~\citen{kozlov1965} and \citen{jerome1967}.
We will introduce the phase diagram expected from this gap equation in Sect.~\ref{sec:BCSBEC}. 

In the above discussion, we assume that the VB top and CB bottom are located at the same $\bm{k}$ point, as shown in Fig.~\ref{fig1} (left). 
However, if the CB bottom is separated from the VB top with a wave vector $\bm{Q}$ in the Brillouin Zone (BZ), the order parameter is characterized by the pair of $ \hat{a}^{\dag}_{\bm{k}+\bm{Q}}$ and $\hat{b}_{\bm{k}}$. 
For example, when $t^{(a)} t^{(b)}>0$ in the 2D square lattice, the VB top and CB bottom located at different $\bm{k}$ points are connected by $\bm{Q}=(\pi,\pi)$.  
When $\bm{Q}\ne \bm{0}$, the order parameter modulates the particle density in real space with the period characterized by $\bm{Q}$. 
Hence, an excitonic order with the modulation vector $\bm{Q}$ gives rise to a density-wave state~\cite{halperin1968,halperin1968_2}. 

At the end of this section, we comment on the terminology. 
Although ``excitonic insulator'' includes ``insulator'' in its name, excitonic orders are not always insulating.  
For example, if there are three bands, such as one VB and two CBs, on the Fermi level, two of the three bands form an excitonic order but the nonbonding band (which is irrelevant to the order parameter) can remain on the Fermi level~\cite{halperin1968_2}.  
In this case, the single-particle band structure is still metallic, and  ``insulator'' is not a reasonable word to represent this metallic state. 
Hence, we call the order (or phase) characterized by the order parameter of electron--hole pairs ``excitonic order'' (or ``excitonic phase'') regardless of whether the band structure is insulating or metallic. 
If the band structure is insulating, we call it ``excitonic insulator.''

\subsection{Phase diagram and BCS--BEC crossover} \label{sec:BCSBEC}

The excitonic order in Sect.~\ref{sec:EO} is characterized by the electron--hole pairs directly driven by the Coulomb interaction. 
Although the components of a pair and driving force differ from the conventional superconducting order, theoretical frameworks such as the gap equation [Eq.~(\ref{eq:Gapeq_Vk})] are analogous to the mathematical structures describing ordered states of paired fermions. 
Hence, the formation of the excitonic order is often discussed as exciton condensation. 
From the perspective of pair condensation, the order in the weak-coupling ($V \ll 1$) regime is BCS-like, where loosely bound pairs form the order.  
On the other hand, in the strong-coupling ($V \gg 1$) regime, tightly bound pairs can exist at high temperatures, and the BEC of these pairs forms the order. 
A smooth crossover connects these two different regimes, and the concept of the BCS--BEC crossover~\cite{eagles1969,leggett1980,nozieres1985,randeria2014,strinati2018} is often applied to the physics of the EI. 

\begin{figure}[b]
\begin{center}  
\includegraphics[width=0.85\columnwidth]{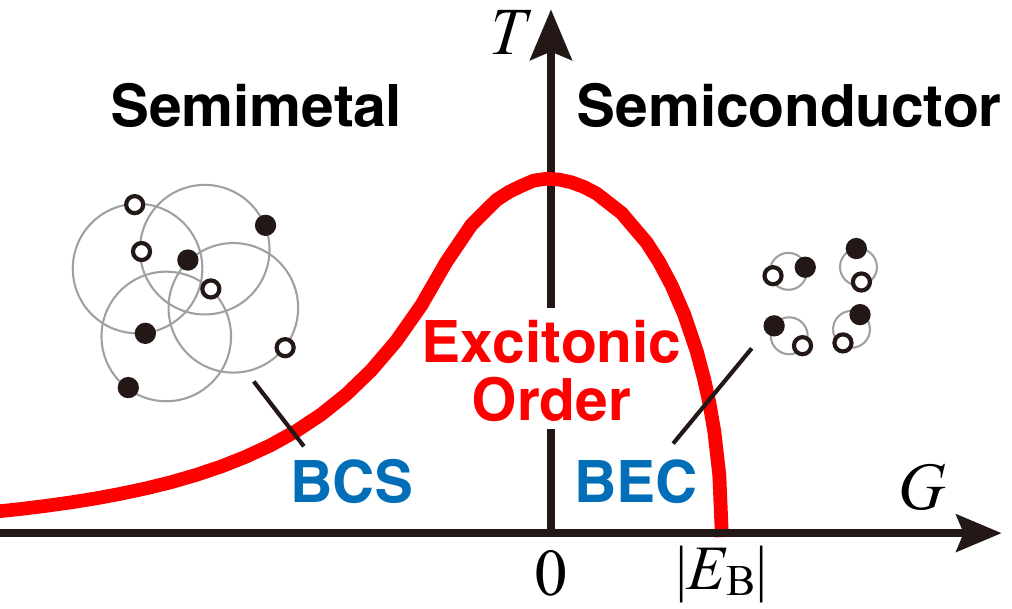}          
\caption{(Color online) Phase diagram of the excitonic order.}
\label{fig2}
\end{center}
\end{figure}

Figure~\ref{fig2} shows the phase diagram suggested in the 1960s~\cite{kozlov1965,jerome1967}. 
The horizontal axis $G$ in Fig.~\ref{fig2} indicates the energy gap between the VB top and CB bottom in the normal state ($\Delta=0$), where $G>0$ corresponds to the semiconductor and $G<0$ corresponds to the semimetal. 
The phase diagram in Fig.~\ref{fig2} is predicted from the gap equation [Eq.~(\ref{eq:Gapeq_Vk})] assuming that the nonlocal Coulomb interaction $V(\bm{q})$ is screened depending on the carrier density~\cite{kozlov1965,jerome1967,bronold2006}.  
The analytical solutions of the gap equation incorporating the screening effects have been provided, such as in Ref.~\citen{kozlov1965}. 

When the VB and CB are deeply overlapped at $G<0$, there are many carriers and the Coulomb interaction is strongly screened. 
When the Coulomb interaction is weak in this semimetallic regime, the transition to the excitonic phase is BCS-like, where the pair formations occur at the transition temperature.  
Similarly to the Cooper instability in the BCS  superconductors~\cite{cooper1956,bardeen1957}, if the Fermi surfaces of the VB and CB are well overlapped (or the nesting of the Fermi surfaces is nearly perfect), a weak interaction can induce instability in the electronic structure around the Fermi level.  
Hence, although the driving interaction is weak, the instability around the Fermi surface can induce the excitonic order. 
In other words, the excitonic order is suppressed when the Fermi surfaces of the VB and CB are anisotropic and their overlap (or nesting) is imperfect~\cite{zittartz1967}.  
Although the phase diagram in Fig.~\ref{fig2} is drawn as if the excitonic phase continues to $G \rightarrow - \infty$, the actual phase boundary in the $G<0$ region is cut off at a finite $|G|$ due to imperfect Fermi-surface overlap in real materials~\cite{zittartz1967}. 

On the other hand, the Coulomb screening in the semiconducting region at $G>0$ is not as strong as in the semimetallic case.  
If the gap $G$ is large, the energy spectrum of the normal semiconductor often contains the peak structure due to the bound state of the electron and hole, i.e., exciton, in the sub-bandgap regime (energy range below $G$). 
In contrast to the BCS case at $G<0$, since the phase transition cannot involve instability caused by the overlap of the Fermi surfaces, the transition requires a Coulomb contribution beyond the energy scale of the band gap $G$.  
As shown in the early stage of studies~\cite{kozlov1965,jerome1967}, the excitonic instability in the semiconductor region occurs when the band gap $G$ is smaller than the magnitude of the exciton binding energy $E_{\rm B}$~\cite{kozlov1965,jerome1967}.   
Hence, as shown in Fig.~\ref{fig2}, the semiconductor turns into the EI at $G\sim |E_{\rm B}|$. 
We can interpret this phase transition as the softening of the exciton mode in the semiconductor (see Sect.~\ref{sec:CM_EI}). 
In the strongly correlated regime, pairs can exist even above the transition temperature, and the ordering can be regarded as the condensation of the pairs.  
In analogy with the physics of the BCS--BEC crossover, the BEC picture is often applied to the phase transition in the semiconductor side~\cite{bronold2006}.   
Note that because many-body effects become strong from the intermediate to the strong-coupling region, theoretical treatment beyond the Hartree--Fock approximation is required in precise quantitative evaluations of crossover phenomena.   

As introduced in this section, the order formations in the weak-coupling (semimetal) and strong-coupling (semiconductor) regions have different characters. 
These two regions are connected via crossover~\cite{bronold2006}, and the idea of the BCS--BEC crossover is often employed to describe the physics of the EI.

\subsection{Perspective as a correlated electron system}

The EI is one of the insulators driven by the Coulomb interaction. 
In correlated electron systems, the Mott insulator (MI) is the first thing that comes to mind as a correlation-driven insulator.
In the MI, half-filled metallic bands split into upper and lower Hubbard bands owing to the on-site Coulomb interaction.  
On the other hand, in the EI, one of the two bands is nearly fully filled and the Coulomb interaction between the VB and CB induces the recomposed insulating state with a gap opening.  
In this light, the EI can be regarded as a correlation-driven insulator realized with different band fillings. 
Here, we consider the EI from the perspective of correlated electron systems. 

\begin{figure}[t]
\begin{center}  
\includegraphics[width=\columnwidth]{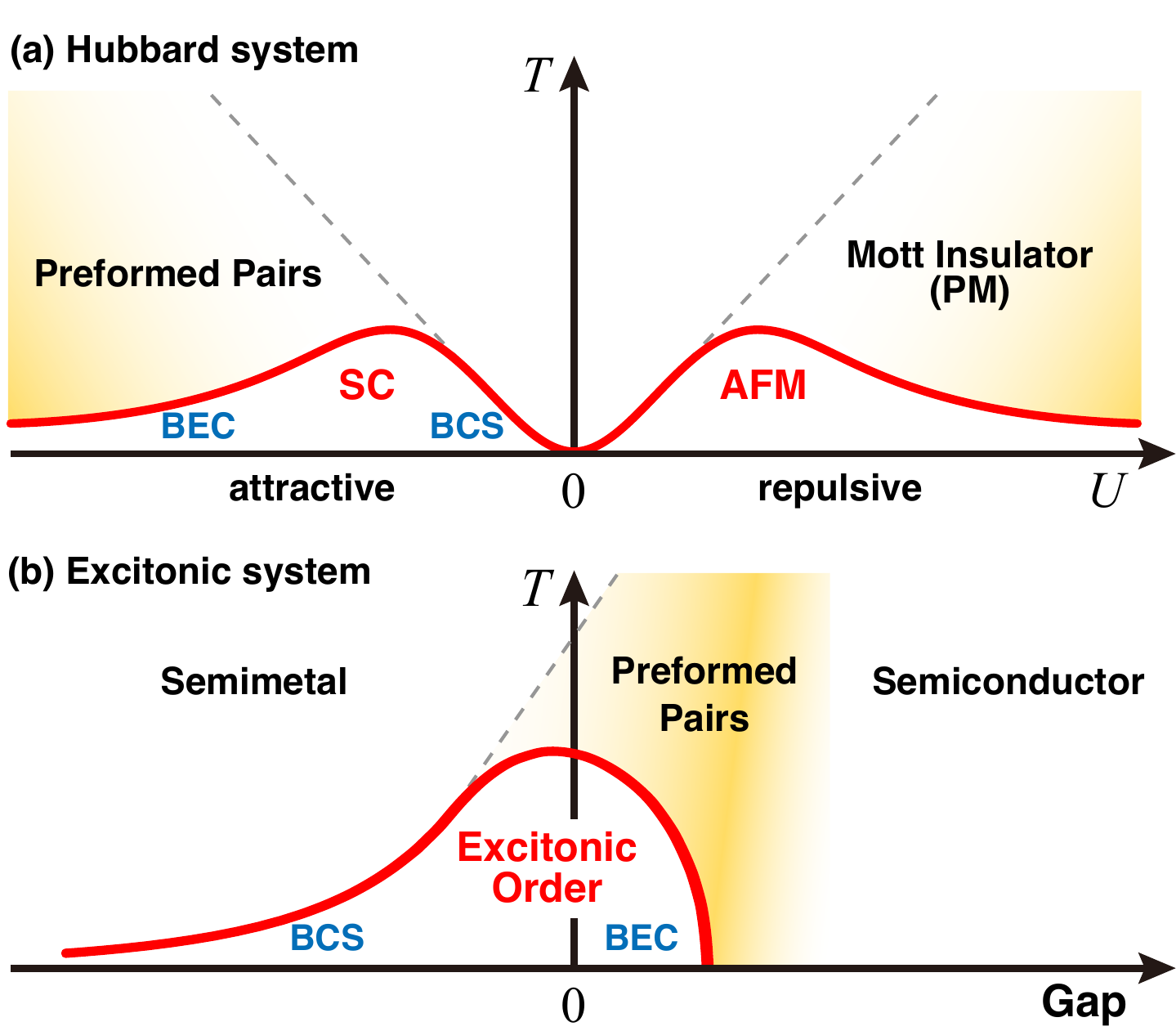}          
\caption{(Color online)
(a) Typical phase diagram in the Hubbard system as a function of the local Coulomb interaction $U$, where red solid and gray dashed lines correspond to $T_{\rm c}$ and $T^*$, respectively. 
(b) Phase diagram in the excitonic system.} 
\label{fig3}
\end{center}
\end{figure}

In Fig.~\ref{fig3}(a), we plot a typical phase diagram of the Hubbard system as a function of the local Coulomb interaction $U$~\cite{esslinger2010,tarruell2018}. 
In the weak-coupling (or BCS) regime, simple metallic states transform into the antiferromagnetic (AFM) state with the repulsive interaction ($U>0$) or the superconducting (SC) state (or charge density wave state) with the attractive interaction ($U<0$) at low temperatures. 
The mean-field theory involving the instability around the Fermi level can be a good approximation in describing these phase transitions in this weak-coupling regime.  
On the other hand, in the strong-coupling regime with $U>0$, the temperature of the gap opening (Mott transition) ($T^*$) and the transition temperature to the AFM ($T_{\rm c}$) do not coincide. 
The paramagnetic (PM) MI state exists in a wide range of temperatures above the AFM state. 
The ordering temperature $T_{\rm c}$ of the spin degrees of freedom is characterized by the spin interaction $J = 4 t_{\rm hop}^2 / U$ (where $t_{\rm hop}$ is a hopping integral). 
Corresponding to the repulsive case, the temperature of the pair formation $T^*$ is higher than the transition temperature $T_{\rm c}$ of the SC state in the attractive BEC regime. 
Hence, the preformed pair state exists above the temperature of the BEC.  
The weak-coupling (BCS) and strong-coupling (BEC) regimes show a smooth crossover. 

By comparison with Hubbard systems, we can obtain several perspectives on excitonic systems. 
In the weak-coupling BCS regime, the gap opening simultaneously occurs at the transition temperature.  
In common with both Hubbard and excitonic systems, the phase transitions are induced by involving the instabilities around the Fermi level since only electrons near the Fermi surfaces can contribute to the phase transitions in the weak-coupling regime.     
On the other hand, in the intermediate and strong-coupling BEC regimes, corresponding to the paramagnetic MI at $U>0$, the phase diagram of the excitonic system would also contain a gapped region above the transition temperature [see Fig.~\ref{fig3}(b)].  
The strong Coulomb fluctuation can deform the band structure even without the order parameter as in the MI. 
The gap opening in this region has been demonstrated by numerical calculations in the two-band correlated models~\cite{seki2014,ejima2021}.  
Since the excitonic order is characterized by electron--hole pairs, the gapped state above $T_{\rm c}$ is often called the ``preformed pair state'' in analogy with pair condensation in attractive fermionic systems. 
In a broad sense, we may regard this gapped state above $T_{\rm c}$ as an ``excitonic insulator'' because the deformation of the electronic structure at $T_{\rm c}<T<T^*$ is also attributed to the excitonic (interband Coulomb) interaction.  
If we accept this terminology, the EI corresponds to the MI at $0\le T < T^*$ while the excitonic order corresponds to the AFM order at $0\le T < T_{\rm c}$.  
Meanwhile, when the gap between the VB top and CB bottom is much larger than the energy scale of the exciton binding energy (i.e., wide-bandgap semiconductor), the fluctuations to the excitonic order are negligible, and the bands become the usual semiconductor-like structure without noticeable band deformation.

\subsection{Spin degrees of freedom} \label{sec:EI_spin}

In the above discussions, spin degrees of freedom are not explicitly considered. 
However, an electron has spin degrees of freedom, which is not a negligible factor in real materials. 
When the electron spins are considered, the excitonic order parameter can be defined by $\braket{\hat{a}^{\dag}_{\bm{k}+\bm{Q},\sigma} b_{\bm{k},\sigma'}}$ with $\sigma, \sigma'=\uparrow,\downarrow$.  
The excitonic order with spin degrees of freedom is usually described in terms of the spin-singlet or spin-triplet order parameters.  
For example, when a local interaction forms an order, the spin-singlet-type order parameter $\phi^{0}_{\bm{Q}}$ is defined as 
\begin{equation}
\phi^{0}_{\bm{Q}} = \frac{1}{N} \sum_{\bm{k}} \sum_{\sigma} \braket{ \hat{a}^{\dag}_{\bm{k}+\bm{Q},\sigma} \hat{b}_{\bm{k},\sigma} },
\label{eq:EIOP_SS}
\end{equation}
whereas the spin-triplet-type order parameters $\bm{\phi}_{\bm{Q}}=(\phi^{x}_{\bm{Q}},\phi^{y}_{\bm{Q}},\phi^{z}_{\bm{Q}})$ are defined as 
\begin{equation}
\bm{\phi}_{\bm{Q}} = \frac{1}{N} \sum_{\bm{k}} \sum_{\sigma,\sigma'} \bm{\sigma}_{\sigma\sigma'}  \braket{ \hat{a}^{\dag}_{\bm{k}+\bm{Q},\sigma} \hat{b}_{\bm{k},\sigma'} }, 
\label{eq:EIOP_ST}
\end{equation}
where $\bm{\sigma}=(\sigma^{x},\sigma^y,\sigma^z)$ is the vector of the Pauli matrices. 
When the driving interaction is nonlocal (or an order parameter is formed across the unit cells), the singlet and triplet order parameters should be defined as in Eq.~(\ref{eq:EIOP_Vk}).  

In the early stage of studies, Halperin and Rice classified excitonic orders with spin degrees of freedom into four categories: (I) singlet and real order parameter modifying the charge density, (II) triplet and real order parameter modifying the spin density, (III) singlet and imaginary order parameter modifying the charge current density, and (IV) triplet and imaginary order parameter modifying the spin current density~\cite{halperin1968,halperin1968_2}.  
Note that this is the classification without spin-orbit coupling~\cite{halperin1968}. 
When the VB top and CB bottom are separated by a vector $\bm{Q}$ in the BZ, the real singlet (triplet) order parameter generates the oscillation of the charge (spin) density in real space with the modulation vector $\bm{Q}$. 
The energies of the singlet and triplet states are the same when the dominant term of the Coulomb interaction is only considered for the order formations~\cite{halperin1968,halperin1968_2}. 
However, additional factors such as exchange interactions and electron--lattice coupling lift the energy degeneracy.   
Halperin and Rice~\cite{halperin1968,halperin1968_2} showed that the lowest energy state in the purely electronic model (including the exchange interactions) is the triplet state (II) with the spin-density modulation when the energy splitting terms are treated as a small perturbation within the Hartree--Fock theory. 
They also mentioned that the singlet state (I) associated with charge density redistribution can have a lower energy if an electron--phonon interaction is sufficiently strong~\cite{halperin1968,halperin1968_2}. 
Although the classes (III) and (IV) are not simply realized~\cite{halperin1968_2}, the order with the imaginary part associated with the charge (spin) current density may correspond to the charge (spin) loop current order in recent terminology~\cite{christensen2022,tazai2023}.   

The actual picture of an excitonic order in real space is not as simple as a conventional density wave characterized by net magnetization or net charge density in a single unit. 
This is because an electron density distribution modulated by excitonic ordering strongly depends on the orbital textures of the VB and CB. 
We will briefly discuss this issue in Sect.~\ref{sec:EI_TB} employing the tight-binding picture.


\section{Excitonic Insulators in Strongly Correlated Systems} \label{sec:EI_SCES}

Various numerical methods have been developed to research strongly correlated electron systems.
In the recent two decades, numerical techniques that can incorporate correlation effects precisely, such as DMFT~\cite{georges1996} and DMRG~\cite{white1992}, have been applied to solve excitonic problems on lattice models~\cite{kunes2015}.  
Such progress is important since precise treatments of the many-body correlations are crucial to investigating the EI states in the wide parameter range up to the strong-coupling BEC regime. 
In this section, as a recent development, we overview the theoretical studies set in strongly correlated lattice models.

\subsection{Spinless model} \label{sec:EI_SCES_spinless}

First, we introduce the EI state realized in the spinless correlated model. 
The Hamiltonian of the spinless two-orbital model reads
\begin{align}
\hat{\mathcal{H}}
=&-t_a \sum_{\langle i,j \rangle} \left( \hat{a}^{\dag}_{i} \hat{a}_{j} +  {\rm H.c.} \right) 
-t_b \sum_{\langle i,j \rangle} \left( \hat{b}^{\dag}_{i} \hat{b}_{j} +  {\rm H.c.} \right) 
\notag \\
&+\frac{D}{2} \sum_{j} \left(\hat{n}_{j,a} - \hat{n}_{j,b} \right) + U \sum_{j} \hat{n}_{j,a} \hat{n}_{j,b} , 
\label{eq:ham_EFKM}
\end{align}
where $\hat{a}_{j}$ ($\hat{a}^{\dag}_{j}$) and $\hat{b}_{j}$ ($\hat{b}^{\dag}_{j}$) are the annihilation (creation) operators of fermions at site $j$ on the $a$ and $b$ orbitals, respectively.
$\hat{n}_{j,a}=\hat{a}^{\dag}_{i} \hat{a}_{j} $ ($\hat{n}_{j,b}=\hat{b}^{\dag}_{i} \hat{b}_{j} $) is the number operator of orbital $a$ ($b$). 
$\langle i, j \rangle$ represents a pair of nearest-neighbor sites. 
$t_a$ ($t_b$) is the hopping integral of a fermion on the $a$ ($b$) orbital network, and $D$ is the energy-level difference between two orbitals. 
$U$ is the interorbital Coulomb interaction. 
We only consider the nearest-neighbor hoppings, whereas this Hamiltonian is essentially the same as $\hat{\mathcal{H}}_0+\hat{\mathcal{H}}_V$ in Eqs.~(\ref{eq:ham_twoband_0}) and (\ref{eq:ham_twoband_V}). 
Although $U$ is the same as $V$ in Eq.~(\ref{eq:ham_twoband_V}), in this section, we use $U$ following the convention of previous studies. 
Here, we introduce the character of the model at $n_a + n_b = 1$ (i.e., one of the two orbitals is occupied), where $n_{\gamma} = (1/L) \sum_j \braket{\hat{n}_{j,\gamma}}$ in the $L$ site system. 

This model is often called the extended Falicov--Kimball model (EFKM). 
The usual Falicov--Kimball model describes correlations between itinerant and localized fermions~\cite{falicov1969,freericks2003}, where $c$ and $f$ are often used for denoting itinerant and localized particles, respectively. 
The extended model allows the hopping between the localized ($f$) orbitals, where the two dispersive bands are realized. 
Since two bands are correlated via the interband interaction $U$, this spinless EFKM is used as a minimal correlated model for the EI~\cite{batista2004,farkaovsky2008,schneider2008,ihle2008,zenker2010,phan2010,phan2011,zenker2011,seki2011,zenker2012,kaneko2013ed,ejima2014,farkasovsky2017,hamada2017,kadosawa2020,mkaneko2021,eaton2024}. 

If we employ the pseudospin representation $(a,b) \rightarrow (\uparrow,\downarrow)$, the Hamiltonian of Eq.~(\ref{eq:ham_EFKM}) corresponds to the Hubbard model with the spin-dependent hopping $t_{\sigma}$ ($=t_a, t_b$) under the magnetic field $h_z$ (=$D$)~\cite{batista2002}. 
In the strong-coupling limit $U \gg t_{\sigma}$, the EFKM at $n_a + n_b = 1$ can be mapped onto the spin-1/2 XXZ Heisenberg model $\hat{\mathcal{H}}_{\rm XXZ}=\sum_{\langle i,j\rangle} \left[ J_{xy} \left( \hat{S}^x_i \hat{S}^x_j + \hat{S}^y_i \hat{S}^y_j \right) + J_z \hat{S}^z_i \hat{S}^z_j \right]$~\cite{batista2002}.   
The spin anisotropy ($J_z/J_{xy}$) in this effective XXZ model originates from the difference in the hopping integrals $t_a \ne t_b$ in the EFKM (see, for example, Ref.~\citen{batista2002} for details).    
The XXZ model (on simple bipartite lattices) has the N\'eel AFM phase due to the Ising anisotropy ($J_{z}/J_{xy}>1$ at $h_z=0$), XY phase due to the in-plane anisotropy ($J_{z}/J_{xy}<1$ at $h_z=0$), and fully spin-polarized ferromagnetic (FM) phase due to a large magnetic field $h_z$. 
These FM, AFM, and XY phases in the effective XXZ model correspond to the band insulator (BI),  staggered orbital order (SOO), and EI phases in the original EFKM~\cite{batista2002}.      
In the SOO phase, fermions occupy the A and B sites alternately on a bipartite lattice.   

\begin{figure}[t]
\begin{center}  
\includegraphics[width=0.85\columnwidth]{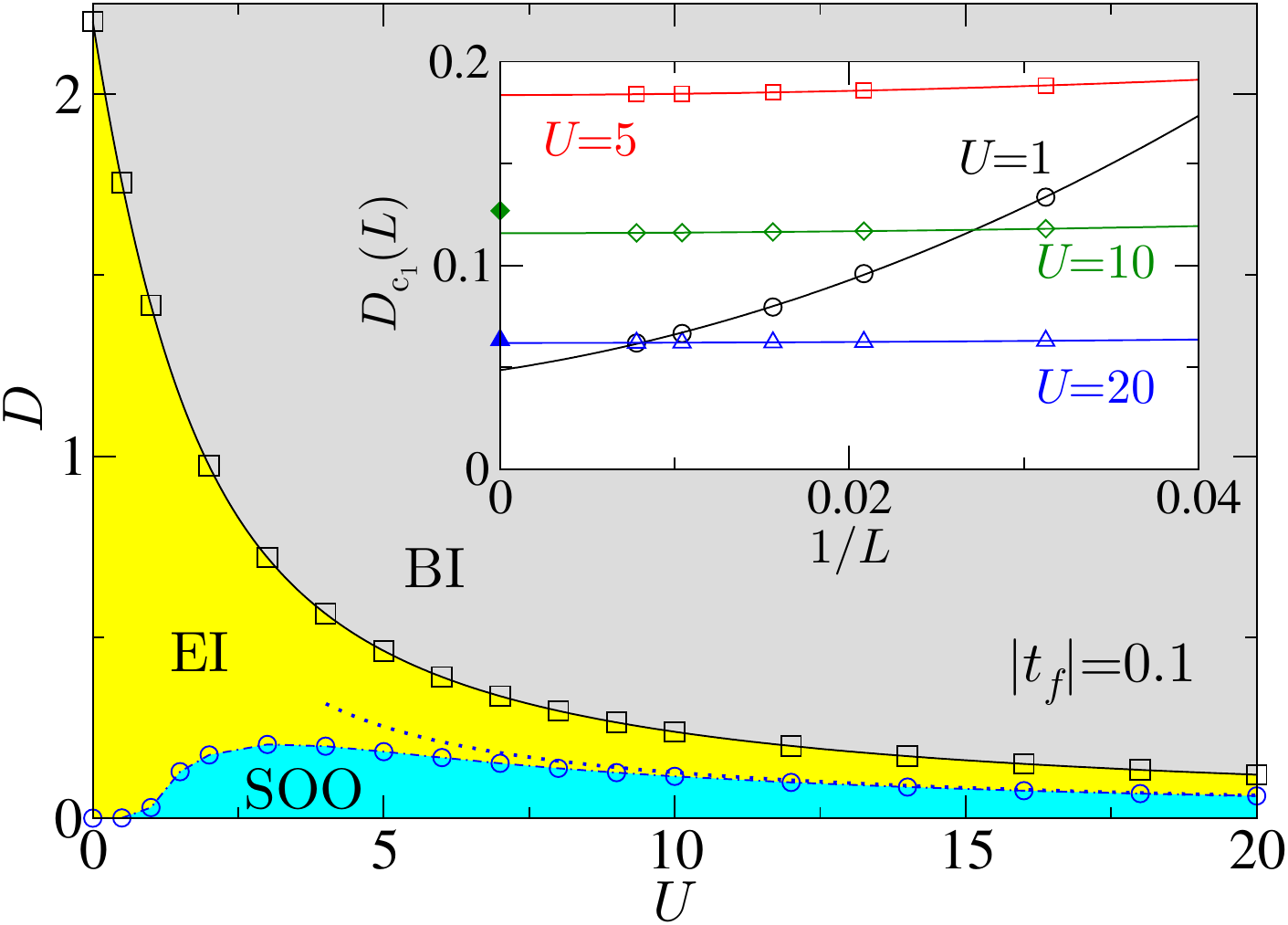}        
\caption{(Color online) 
Ground-state phase diagram of the 1D EFKM with $t_b = -0.1$ ($=t_f$) evaluated by the DMRG method, where $t_a=1$ is the unit of energy. 
BI, EI, and SOO denote band insulator, excitonic insulator, and staggered orbital order, respectively.
The inset shows the finite-size scaling of the phase boundary $D_{c_1}$ between the EI and SOO. 
Reproduced from Ref.~\citen{ejima2014} \copyright $\,$ 2014 American Physical Society.} 
\label{fig4}
\end{center}
\end{figure}

Figure~\ref{fig4} shows the ground-state ($T=0$) phase diagram of the one-dimensional (1D) EFKM evaluated by the DMRG method~\cite{ejima2014}.  
Since the DMRG can reach the unbiased numerical solution in 1D systems, Fig.~\ref{fig4} can be regarded as the exact phase diagram of the 1D EFKM. 
As anticipated by the effective XXZ model, the phase diagram of the 1D EFKM comprises three phases: (I) BI phase in which the low-energy orbital is fully occupied, i.e., $n_b=1$ and $n_a=0$, (II) SOO phase characterized by the long-range correlation of $\braket{( \hat{n}_{j+x,a} - \hat{n}_{j+x,b} ) ( \hat{n}_{j,a} - \hat{n}_{j,b} )} \propto (-1)^x$, and (III) EI phase characterized by the development of an excitonic correlation function $\braket{\hat{a}^{\dag}_{j+x} \hat{b}_{j+x} \hat{b}^{\dag}_j \hat{a}_j }$.  
When $D \gg t_{\gamma}$, the low-energy orbital is fully occupied, i.e., the ground state is the BI.  
A large $U$ also prefers to be the BI. 
On the other hand, when $D\sim 0$, the occupancies of two orbitals are nearly equal, i.e., $n_a \sim n_b \sim 0.5$. 
The SOO phase appears if $|t_b| \ne t_a$, whereas the SOO and EI states are degenerate at $|t_b| = t_a$ and $D=0$, which corresponds to the Heisenberg point in the effective model~\cite{batista2002,batista2004}.   
The EI phase exists in the intermediate orbital filling $0<n_b-n_a< 1$ between the BI and SOO phases. 
As in the gapless XY phase in the 1D XXZ model~\cite{takahashi1999}, the EI phase of the 1D EFKM is a critical phase in which the central charge of the conformal field theory is $c=1$~\cite{ejima2014}.  
The DMRG calculations also show that the excitonic correlation and binding energy are enhanced as $U$ increases corresponding to the crossover between the BCS and BEC regimes~\cite{ejima2014}. 

The EFKMs in $d \ge 2$ dimensional systems have been investigated by various methods, such as Hartree--Fock approximation (HFA)~\cite{farkaovsky2008,schneider2008,ihle2008},  constrained path Monte-Carlo (CPMC) method~\cite{batista2004}, and variational cluster approach (VCA) based on exact diagonalization (ED)~\cite{seki2011}.  
When $d\ge 2$, the critical EI phase in the 1D model is replaced by the ordered excitonic phase (at $T=0$)~\cite{batista2004}.   
As shown in Ref.~\citen{farkaovsky2008}, the phase boundaries between the ground-state ($T=0$) phases in the 2D EFKM obtained by the HFA show good agreement with those calculated by the CPMP method (when $U/t_a=2$)~\cite{batista2004}.  
The phase diagram at $T>0$ has been plotted by the HFA and slave-boson approach~\cite{zenker2010}. Although both methods qualitatively reproduce a dome-like phase diagram similar to Fig.~\ref{fig3}(b), the transition temperature to the excitonic order evaluated by the slave-boson technique is lower than that by the HFA, implying the importance of the correlation effects at nonzero temperatures~\cite{zenker2010}. 

Although the spinless EFKM is the simplest model for the EI, the model can describe essential aspects of excitonic physics including the BCS--BEC crossover. 
For example, the BCS--BEC crossover of the EFKM has been examined using the anomalous momentum distribution function associated with the excitonic order parameter~\cite{phan2010,seki2011}. 
When spin degrees of freedom are not crucial in a target system, the spinless EFKM can be a reasonable model for describing an EI state. 
For example, if a candidate material is discussed as a spin-singlet EI that can be stabilized by electron--phonon interactions, an EFKM coupled with lattice degrees of freedom gives a minimum model for phonon-coupled EI states~\cite{zenker2014,do2017,murakami2017,do2022}.   
Since we can suppress numerical costs of the fermionic part compared with models with spin degrees of freedom, the EFKM may enable us to obtain numerically precise results (e.g., with larger system size and small truncation error).

\subsection{Model with spin degrees of freedom} \label{sec:EI_SCES_spin}

Next, we consider a model with spin degrees of freedom.  
A simple correlated model with spin degrees of freedom is the two-orbital Hubbard model (TOHM), whose Hamiltonian is given by 
\begin{align}
\hat{\mathcal{H}}
=&-t_a \sum_{\langle i,j \rangle} \sum_{\sigma} \left( \hat{a}^{\dag}_{i,\sigma} \hat{a}_{j,\sigma} +  {\rm H.c.} \right) 
-t_b \sum_{\langle i,j \rangle} \sum_{\sigma} \left( \hat{b}^{\dag}_{i,\sigma} \hat{b}_{j,\sigma} +  {\rm H.c.} \right) 
\notag \\
&+\frac{D}{2} \sum_{j} \left(\hat{n}_{j,a} - \hat{n}_{j,b} \right) 
+ U \sum_{j} \left( \hat{n}_{j,a,\uparrow} \hat{n}_{j,a,\downarrow} + \hat{n}_{j,b,\uparrow} \hat{n}_{j,b,\downarrow} \right)
\notag \\
&+ U' \sum_{j} \hat{n}_{j,a}  \hat{n}_{j,b}
 -2 J \sum_{j} \left(  \hat{\bm{S}}_{j,a}  \cdot \hat{\bm{S}}_{j,b} + \frac{1}{4} \hat{n}_{j,a}  \hat{n}_{j,b} \right)
 \notag \\
& + J' \sum_{j} \left( \hat{a}^{\dag}_{j,\uparrow} \hat{a}^{\dag}_{j,\downarrow} \hat{b}_{j,\downarrow} \hat{b}_{j,\uparrow} +  {\rm H.c.} \right) . 
 \label{eq:ham_TOHM}
\end{align}
$\hat{a}_{j,\sigma}$ ($\hat{a}^{\dag}_{j,\sigma}$) and $\hat{b}_{j,\sigma}$ ($\hat{b}^{\dag}_{j,\sigma}$) are the annihilation (creation) operators of fermions with spin $\sigma = \uparrow, \downarrow$ at site $j$ on the $a$ and $b$ orbitals, respectively. 
$\hat{n}_{j,a,\sigma}=\hat{a}^{\dag}_{j,\sigma} \hat{a}_{j,\sigma}$ ($\hat{n}_{j,b,\sigma}=\hat{b}^{\dag}_{j,\sigma} \hat{b}_{j,\sigma}$) is the number operator of orbital $a$ ($b$), and $\hat{n}_{j,\gamma}=\sum_{\sigma} \hat{n}_{j,\gamma,\sigma}$ ($\gamma = a,b$). 
$\hat{\bm{S}}_{j,a}= (1/2) \sum_{\sigma,\sigma'} \hat{a}^{\dag}_{j,\sigma} \bm{\sigma}_{\sigma\sigma'} \hat{a}_{j,\sigma'}$ ($\hat{\bm{S}}_{j,b}= (1/2) \sum_{\sigma,\sigma'} \hat{b}^{\dag}_{j,\sigma} \bm{\sigma}_{\sigma\sigma'} \hat{b}_{j,\sigma'}$) is the spin operator of orbital $a$ ($b$), where $\bm{\sigma}=(\sigma^x,\sigma^y,\sigma^z)$ is the vector of the Pauli matrices. 
As in the EFKM, $t_a$ ($t_b$) is the hopping integral on the $a$ ($b$) orbital network, and $D$ is the energy-level difference between two orbitals. 
$U$ is the on-site Coulomb interaction while $U'$ is the interorbital Coulomb interaction. 
Note that $U'$ in the TOHM corresponds to the interorbital interaction $U$ in the EFKM. 
$J$ corresponds to Hund's coupling, and $J'$ is the strength of the pair hopping. 
Usually, $J' = J$. 
Here, we consider the case at $n_a + n_b = 2$ in the two-orbital system, where the density of orbital $\gamma$ ($=a,b$) is defined as $n_{\gamma}=(1/L) \sum_{j,\sigma} \braket{\hat{n}_{j,\gamma,\sigma}}$. 

\begin{figure}[t]
\begin{center}  
\includegraphics[width=0.95\columnwidth]{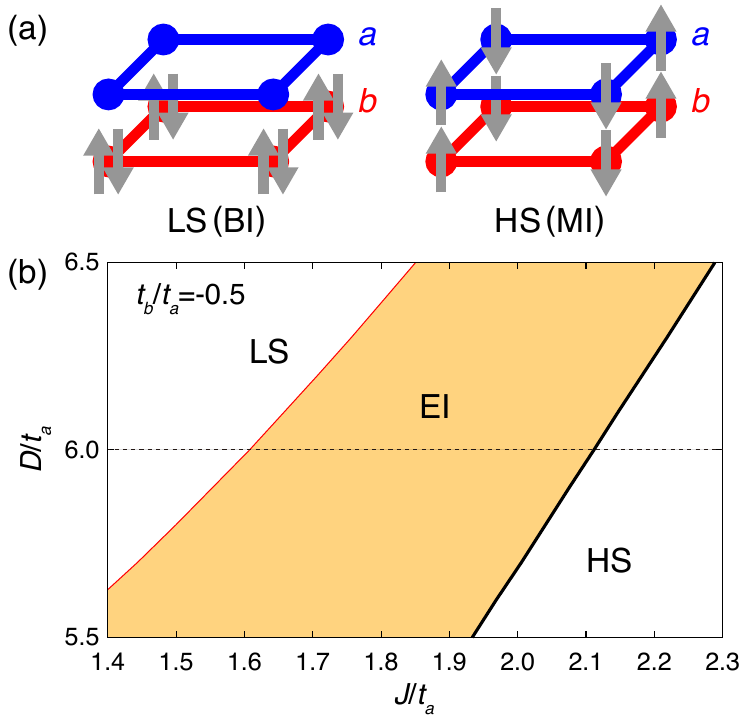}          
\caption{(Color online) 
(a) Schematic figures of the low-spin (LS) (left) and high-spin (HS) (right) states. 
The LS and HS states correspond to the band insulator (BI) and Mott insulator (MI) states, respectively.
(b) Ground-state phase diagram of the effective model for the TOHM evaluated by the mean-field theory, where $U=6J$, $U'=4J$, $J'=J$, and $t_b/t_a=-0.5$ were used. 
EI denotes the excitonic insulator phase. 
The figure (b) is reproduced from Ref.~\citen{nasu2016} \copyright $\,$ 2016 American Physical Society.} 
\label{fig5}
\end{center}
\end{figure}

There are two trivial insulating phases in the TOHM~\cite{werner2007,suzuki2009,kunes2011,higashiyama2008}.   
When $D \gg U$, the lower-energy orbitals are fully occupied ($n_b=2$) while the other orbitals are empty ($n_a=0$).  
This corresponds to the BI phase [see Fig.~\ref{fig5}(a)]. 
On the other hand, when $U \gg D$, both orbitals are half-filled ($n_a=n_b=1$) and all sites are singly occupied to avoid the energy loss of $U$. 
This insulating phase caused by $U$ is the MI phase [see Fig.~\ref{fig5}(a)]. 
Hence, the BI and MI phases reside in the limits $D \gg U$  and $U\gg D$, respectively. 
The TOHM has been employed to describe the spin-state crossover, such as in the cobalt oxides~\cite{suzuki2009,kunes2011}.   
In the context of the spin-state crossover, the BI corresponds to the low-spin (LS) state whereas the MI with a strong $J$ corresponds to the high-spin (HS) state, where the competing $D$ and $J$ give rise to complex many-body problems.  
These LS (BI) and HS (MI) states are schematically shown in Fig.~\ref{fig5}(a).  

The EI phase can emerge between the BI and MI phases because the semimetallic band overlap at $0 < n_a < n_b <2$ can induce the excitonic instability between the VB and CB via the interorbital Coulomb interactions~\cite{balents2000prb}.  
Indeed, the presence of the EI phase in the intermediate regime has been verified by various numerical methods, such as HFA~\cite{brydon2009_1,brydon2009_2,zocher2011,koga2024}, VCA~\cite{kaneko2012,kaneko2014,kaneko2015,fujiuchi2018}, DMFT~\cite{kunes2014,kunes2014_2,niyazi2020}, QMC~\cite{huang2022}, and DMRG~\cite{kitagawa2022}. 
The effective spin/pseudospin model has been derived from the strong-coupling expansion, and the emergence of the EI phase and its physical properties have also been studied in the effective model~\cite{nasu2016,nasu2020,nasu2021}. 
When $J=J'=0$, the spin-singlet and spin-triplet excitonic states formed by $U'$ are degenerate corresponding to the early assessment by Halperin and Rice~\cite{halperin1968}.  
In the TOHM, Hund's coupling term lifts the degeneracy~\cite{brydon2009_2,kaneko2014,kunes2014}.  
With respect to the singlet and triplet excitonic pair operators $\hat{p}^{{\rm 0}\dag}_j= (1/\sqrt{2}) \sum_{\sigma} \hat{a}^{\dag}_{j,\sigma} \hat{b}_{j,\sigma}$ and $\hat{\bm{p}}^{\dag}_j= (1/\sqrt{2}) \sum_{\sigma,\sigma'} \hat{a}^{\dag}_{j,\sigma} \bm{\sigma}_{\sigma\sigma'} \hat{b}_{j,\sigma'}$, respectively, Hund's coupling term $-2J\hat{\bm{S}}_{j,a} \cdot \hat{\bm{S}}_{j,b}$ can be decomposed as $(3J/2) \hat{p}^{0\dag}_j \hat{p}^0_j - (J/2) \hat{\bm{p}}^{\dag}_j \cdot \hat{\bm{p}}_j$. 
Therefore, the energy of the triplet pair becomes lower than that of the singlet pair, i.e., Hund's coupling stabilizes the spin-triplet excitonic state. 
The pair hopping term $J'$ can support the stabilization of the triplet EI phase~\cite{kaneko2015,koga2024}.
$J'$ fixes the phase $\theta$ of the order parameter because the pair hopping $\hat{a}^{\dag}_{j,\uparrow} \hat{a}^{\dag}_{j,\downarrow} \hat{b}_{j,\downarrow} \hat{b}_{j,\uparrow} +  {\rm H.c.}$ leads to the $\cos 2\theta$ contribution in the energy landscape of the order parameter~\cite{littlewood1996}. 

Figure~\ref{fig5}(b) shows the phase diagram of the effective model for the TOHM on the 2D square lattice evaluated by Nasu {\it et al.}~\cite{nasu2016} 
As in the spin-state crossover system, the LS (BI) phase resides in the large $D$ region whereas the HS (MI) phase is realized in the large $J$ region. 
The EI phase emerges between the LS and HS phases. 
Since the TOHM contains various model parameters such as the hopping ratio $t_b/t_a$ and the interactions $U$, $U'$, and $J$, the ground state can be another ordered state depending on the parameters~\cite{nasu2016,niyazi2020,koga2024}. 
For example, the ground-state phase diagram exhibits the spin-state (LS/HS) ordered phase when $t_b/t_a$ is small~\cite{nasu2016}.  

The Hamiltonian of Eq.~(\ref{eq:ham_TOHM}) does not consider orbital hybridization due to hoppings. 
However, even though two orbitals are orthogonal at a single site, the intersite hopping between different orbitals, so-called cross hopping~\cite{kunes2016}, is possible. 
As suggested in previous studies~\cite{kunes2016,geffroy2018,nishida2019,yamamoto2020,shinjo2023}, the cross hopping $\hat{\mathcal{H}}_X=-\sum_{j,\nu,\sigma} \left( t_{ab}^{(\nu)} \hat{a}^{\dag}_{j+\nu,\sigma} \hat{b}_{j,\sigma} + t_{ba}^{(\nu)}  \hat{b}^{\dag}_{j+\nu,\sigma} \hat{a}_{j,\sigma} +  {\rm H.c.} \right) $ fixes the phase of the order parameter and can generate the imaginary order parameter depending on the symmetry of the hoppings.  
Interestingly, a doped spin-triplet EI exhibits a $\bm{k}$-space spin texture associated with the symmetry of the cross hopping~\cite{kunes2016}.  
For example, the doped triplet EI with the $d$-type hopping symmetry $t^{(x)}_{ab}=t^{(x)}_{ba}=-t^{(y)}_{ab}=-t^{(y)}_{ba}$ in the 2D square lattice gives rise to the $d$-wave-like spin splitting of the bands in the $\bm{k}$-space (see, for example, Ref.~\citen{kunes2016} for details). 

Hund's coupling $J>0$, which leads to the ferromagnetic spin alignments within a single atom, is unfavorable for the spin-singlet EI state.  
However, an electron--lattice interaction that couples to the singlet order parameter, such as $gu_{\bm{Q}} \sum_{\bm {k},\sigma} \hat{a}^{\dag}_{\bm{k}+\bm{Q},\sigma}\hat{b}_{\bm{k},\sigma} + {\rm H.c.}$, can support the stabilization of the singlet excitonic order~\cite{zenker2014,kaneko2015}.  
When the support from the electron--lattice contribution is sufficiently larger than the ferromagnetic contributions, the singlet EI state is selected as the ground state. 
Although Hund's coupling is active when two orbitals are located on the same atom, the ferromagnetic flavor may be weak when the valence and conduction orbitals are located on different atoms. 
In the latter case, the spin-singlet excitonic state can emerge with support from electron--lattice couplings.

\subsection{Tight-binding picture of excitonic orders} \label{sec:EI_TB}

The models introduced in Sects.~\ref{sec:EI_SCES_spinless} and \ref{sec:EI_SCES_spin} are based on the tight-binding picture. 
However, the actual images of excitonic orders in real space are not as simple as those of the conventional charge and magnetic orders.
This is because the excitonic order involves the spontaneous hybridization of different electronic bands, and its real-space picture strongly depends on the characters of the orbitals composing the VB and CB.
Here, we consider two cases based on the tight-binding picture: (I) two orbitals that compose the VB and CB are in the same atom, and (II) two orbitals are in different atoms.

\begin{figure}[t]
\begin{center}  
\includegraphics[width=0.85\columnwidth]{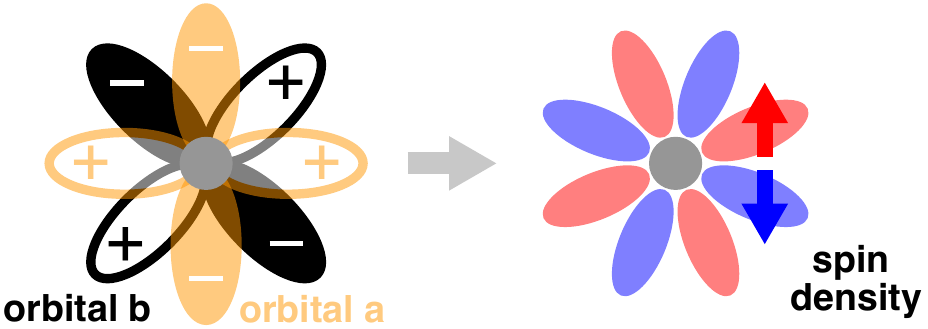}          
\caption{(Color online) 
A schematic figure of the electronic state in a spintriplet excitonic order when the VB and CB are composed of two orbitals in the same atom. 
Two orbitals ($d_{x^2-y^2}$ and $d_{xy}$ orbitals) on the same atom (left) and spin density distribution expected from the hybridized entity due to the spin-triplet excitonic order (right).} 
\label{fig6}
\end{center}
\end{figure}

In case (I), a hybridized entity in the same atom may compose a multipole structure~\cite{kunes2014Co,nasu2016,kaneko2016}. 
For example, when the $d_{xy}$ and $d_{x^2-y2}$ orbitals in the same atom hybridize as a consequence of the triplet excitonic order, two orbitals may create a complicated density distribution in each unit cell~\cite{nasu2016}.  
With the atomic wave functions $\psi_a(\bm{r})$ and $\psi_b(\bm{r})$ for the $a=d_{x^2-y^2}$ and $b=d_{xy}$ orbitals, respectively, the local spin density $s^z(\bm{r})$ in a unit cell is roughly characterized by $\psi_a(\bm{r}) \psi_b(\bm{r}) {\rm Re}[\phi^z_{\rm X}]$ when the spin-triplet-type order parameter $\phi^z_{\rm X}$ ($\propto \sum_{\sigma} \sigma \braket{\hat{a}^{\dag}_{\sigma}\hat{b}_{\sigma}}$) has a nonzero real value. 
If this kind of density distribution is actualized, a spin density with sign changes associated with the product of two different wave functions $\psi_a(\bm{r}) \psi_b(\bm{r})$ appear in a unit cell.    
As schematically shown in Fig.~\ref{fig6}, the hybridized entity may create a magnetic multipole structure with complicated polarization in the local spin density~\cite{nasu2016,kaneko2016}.  
When the density distribution is only characterized by higher-order magnetic multipoles, the total magnetization (magnetic dipole) per single atom is zero. 
In this case, the electronic structure is not simply described as the N\'{e}el AFM order, in which the up and down arrows are alternately placed on the periodic lattice, and the excitonic order rather becomes like a hidden order. 
As we will introduce later, complicated magnetic multiple structures have been predicted in the cobalt-based candidates because the spin-triplet-type order is expected due to strong Hund's coupling~\cite{kunes2014Co,nasu2016,yamaguchi2017}.  

On the other hand, in case (II), a hybridized entity between two different atoms may form an intersite bonding orbital as a consequence of an order formation~\cite{kaneko2016}.   
The orbital bonding across different sites results in a charge deviation, similar to bond-centered charge ordering~\cite{vandenbrink2008}.  
Charge deviations due to bond formation may tend to couple with lattice degrees of freedom. 
The Hund's coupling-like exchange interaction, which is strong among orbitals within the same atom, is less active than in case (I). 
Since electron--phonon interactions can support the stabilization of spin-singlet-type states, a spin-singlet EI associated with lattice distortions tends to arise when two orbitals are in different atoms. 
This may be a reason why the spin-singlet excitonic states are implicitly anticipated in the EI candidates with lattice distortions, such as TiSe$_2$ and Ta$_2$NiSe$_5$ (see Sects.~\ref{sec:tise2} and \ref{sec:tns}).  

As seen from these examples, it is not easy to describe real-space pictures of excitonic orders in one simple manner. 
When we consider real-space images of EIs in the tight-binding picture, it is necessary to consider the shapes of atomic orbitals and their positions. 
Put another way, EIs can be interesting hosts for various unconventional electronic states.


\section{Candidate Materials} \label{sec:candidates}

In this section, we introduce materials that have been discussed in the context of the EI or excitonic ordering. 
In particular, we focus on TiSe$_2$ (Sect.~\ref{sec:tise2}) and Ta$_2$NiSe$_5$ (Sect.~\ref{sec:tns}), which are expected to be the spin-singlet type with lattice distortions, and the Co oxides with the $d^6$ configuration (Sect.~\ref{sec:co_based}), which have been discussed as a spin-triplet-type excitonic order. 
Other candidate materials proposed to date are briefly reviewed in Sect.~\ref{other_candidates}. 
We note that the following sections include topics under debate and potentially contain some statements that may be invalidated by future research.

\subsection{TiSe$_2$} \label{sec:tise2}

\begin{figure}[t]
\begin{center}  
\includegraphics[width=0.95\columnwidth]{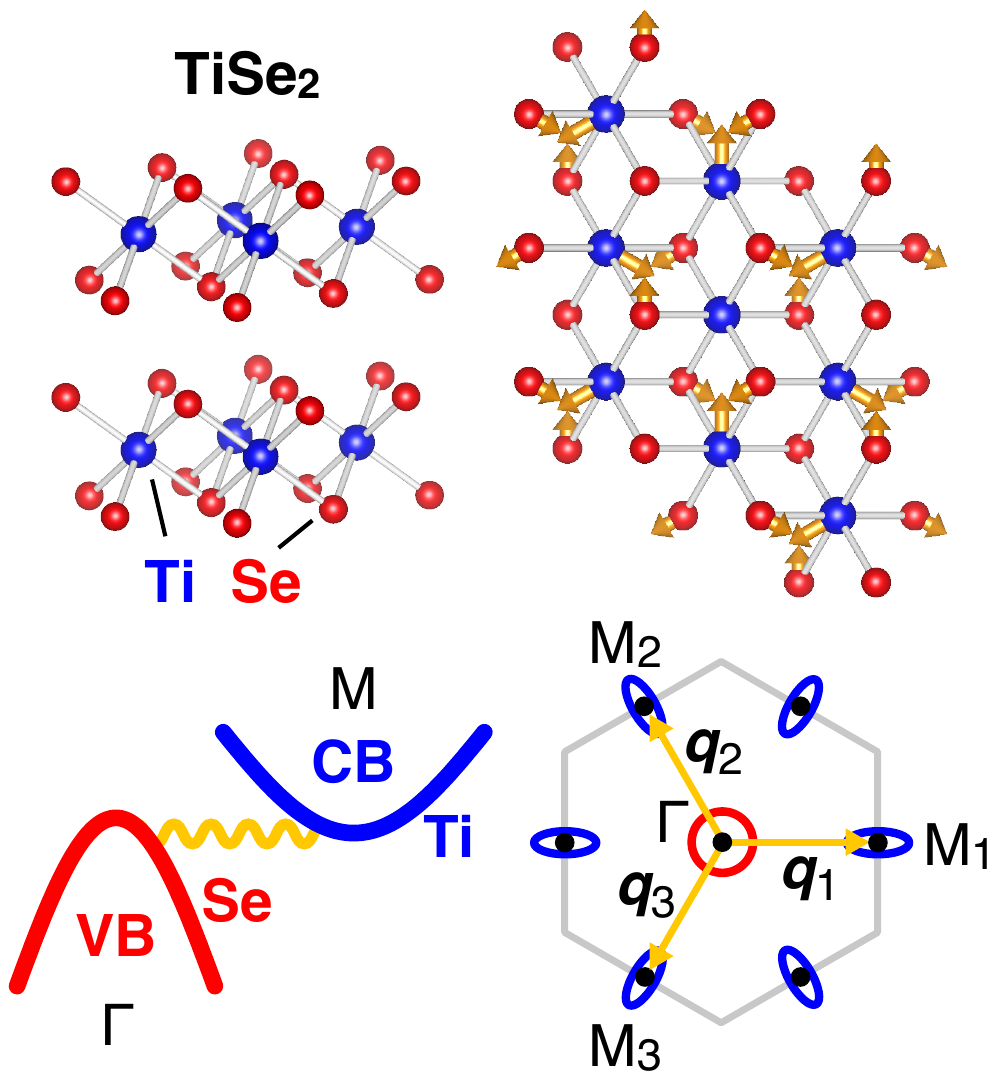}          
\caption{(Color online) 
Crystal structure of 1$T$-TiSe$_2$ (upper left panel) and periodic lattice distortion at low temperatures (upper right panel) visualized by VESTA~\cite{VESTA}, where the arrows indicate the directions of the displacements. 
Schematic band structure and Fermi surfaces of the normal state in TiSe$_2$ (lower panel).} 
\label{fig7}
\end{center}
\end{figure}

The transition-metal dichalcogenide (TMD) 1$T$-TiSe$_2$ is a material with a long history and has often been studied in association with the EI~\cite{traum1978}.    
As in other TMDs with the 1$T$-type structure, the octahedrons, in which six Se atoms are coordinated around a Ti atom, form a single layer (see Fig.~\ref{fig7}). 
The Ti atoms form a triangular lattice in a single layer, and van der Waals layer stacking forms the bulk crystal. 
TiSe$_2$ was synthesized in the 1960s~\cite{greenaway1965}, and its electronic and structural properties have been comprehensively investigated by Di Salvo {\it et al.}~\cite{disalvo1976}. 
The electronic transport properties are metallic, whereas the electronic resistivity once increases around $T= 200$~K reflecting the phase transition~\cite{disalvo1976}.   
Neutron diffraction reveals a second-order structural transition to a $2\times 2 \times 2$ superlattice structure at $T\sim 200$~K~\cite{disalvo1976}. 
Figure~\ref{fig7} (upper right panel) shows the $2\times 2$ periodic lattice distortion in a TiSe$_2$ single layer. 
The X-ray studies also show the phonon softening corresponding to the $2\times 2 \times 2$ superlattice formation~\cite{holt2001,weber2011,kitou2019}.  
Hence, TiSe$_2$ undergoes the phase transition accompanied by the periodic lattice distortion at $T\sim 200$~K. 
This low-temperature phase is referred to as the charge-density-wave (CDW) phase.

TiSe$_2$ is a group-4 TMD~\cite{greenaway1965,chhowalla2013}, whose electron configuration is formally $d^0$. 
Hence, in group-4 TMDs, the transition metal $d$ orbitals mainly compose the CBs and the ligand $p$ orbitals compose the VBs. 
The sulfides TiS$_2$, ZrS$_2$, and HfS$_2$ have semiconducting band structures, in which the VB top and CB bottom are located at the $\Gamma$  and M (L) points, respectively, in the BZ~\cite{zhao2017,glebko2018}.  
However, TiSe$_2$ has a semimetallic band structure, where both valence $p$-based bands and conduction $d$-based bands cross the Fermi level~\cite{zunger1978,fang1997}.  
As schematically shown in Fig.~\ref{fig7} (bottom panel), the hole and electron pockets at the $\Gamma$ and M (L) points are small. 
Since the band overlap is small, TiSe$_2$ possesses a suitable band structure for excitonic ordering. 
This is the reason why TiSe$_2$ has been discussed in the context of the EI. 

\begin{figure}[t]
\begin{center}  
\includegraphics[width=0.9\columnwidth]{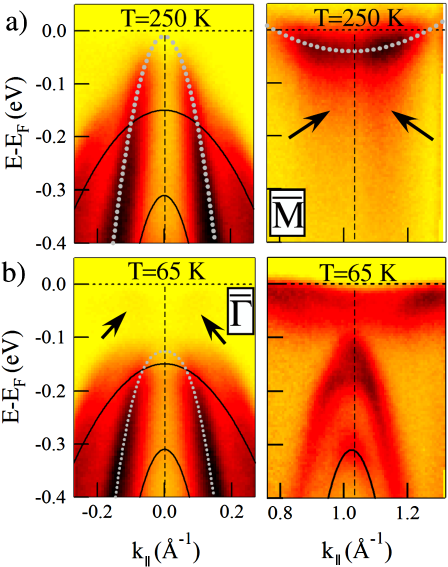}          
\caption{(Color online) 
ARPES spectra of 1$T$-TiSe$_2$ for (a) above and (b) below the transition temperature. 
Reproduced from Ref.~\citen{cercellier2007} \copyright $\,$ 2007 American Physical Society.} 
\label{fig8}
\end{center}
\end{figure}

Angle-resolved photoemission spectroscopy (ARPES) experiments reveal the electronic band structure of TiSe$_2$~\cite{traum1978,pillo2000,kidd2002,cercellier2007,qian2007,chen2015,sugawara2016,watson2019}. 
Figure~\ref{fig8} shows the ARPES spectra of 1$T$-TiSe$_2$ reported by Cercellier {\it et al.}~\cite{cercellier2007}. 
Above the transition temperature $T_{\rm c}$ [see Fig.~\ref{fig8}(a)], the spectra corresponding to the VB top and CB bottom are observed at the $\Gamma$ and M (L) points, respectively, as anticipated from the band calculations.  
On the other hand, at temperatures below $T_{\rm c}$ [see Fig.~\ref{fig8}(b)], the ARPES spectra are modified from the band structure in the normal phase. 
Associated with the 2$\times$2($\times$2) order formation below $T_{\rm c}$, the folding band, which reflects the convex of the VB at the $\Gamma$ point, emerges at the M (L) point. 
In addition, the top of the VB shifts to the lower energy as temperature decreases. 
This energy shift suggests the gap opening due to the reorganization of the energy bands associated with the order formation. 
Similar ARPES spectra have also been observed in the single-layer TiSe$_2$, where the temperature-dependent energy shift is well fitted by the mean-field-like function~\cite{chen2015,sugawara2016}.  
As seen in these studies, the changes in the electronic structure associated with the phase transition have been distinctly observed in the ARPES experiments. 

The controversy in TiSe$_2$ is the origin of the phase transition. 
There are two key contributors to the phase transition: electron--lattice coupling and excitonic interaction. 
The first one is the electron--lattice coupling that leads to the 2$\times$2($\times$2) periodic lattice distortion in TiSe$_2$. 
Note that ($\times$2) indicates the lattice distortion in bulk TiSe$_2$. 
The structural phase transition in TiSe$_2$ was investigated by Motizuki and colleagues in the 1980s~\cite{motizuki1986,yoshida1980,takaoka1980,suzuki1985}.  
Since the CB bottoms are located at three M points, M$_1$, M$_2$, and M$_3$, the phase transition in TiSe$_2$ is characterized by three vectors $\bm{q}_1$, $\bm{q}_2$, and $\bm{q}_3$ connecting the $\Gamma$ point to the M$_1$, M$_2$, and M$_3$ points, respectively (see Fig.~\ref{fig7}).  
Note that the M points are replaced by the L points in bulk TiSe$_2$, and $\bm{q}_j$ ($j=1,2,3$) becomes a three-dimensional vector. 
One of the transverse phonon modes at each M point is unstable in TiSe$_2$, and the transverse modes at three M points simultaneously show softening at $T_{\rm c}$. 
The superposition of the three transverse phonon modes characterized by $\bm{q}_1$, $\bm{q}_2$, and $\bm{q}_3$ gives the low-temperature 2$\times$2($\times$2) lattice structure~\cite{motizuki1986,yoshida1980,takaoka1980,suzuki1985,monney2011}.   
An ordered state in which three $\bm{q}$ modes simultaneously contribute is called the ``triple-$\bm{q}$'' state. 
The phonon softening and the stabilization of the triple-$\bm{q}$ structure have also been demonstrated by the DFT calculations~\cite{calandra2011,bianco2015,douong2015,hellgren2017,singh2017,zhou2020}. 
The 2$\times$2($\times$2) lattice structure resulting from the simultaneous softening of the three phonon modes is consistent with the experimentally observed lattice structure shown in Fig.~\ref{fig7}~\cite{disalvo1976,holt2001,weber2011,kitou2019}.  
Therefore, the electron--lattice interaction is an essential factor for explaining the superlattice formation in TiSe$_2$. 

The second factor is the interband Coulomb interaction that causes the excitonic instability.  
Since TiSe$_2$ has the semimetallic band structure with small hole and electron pockets at $\Gamma$ and M points, an excitonic order characterized by the vector $\bm{q}_j$ potentially develops in the low-temperature state. 
From this perspective, the realizations of an excitonic order in TiSe$_2$ have been theoretically demonstrated~\cite{cercellier2007,monney2009,monney2010,monney2010_NJP,monney2012,monney2015,monney2012_NJP,chen2018,bok2021}. 
Mechanisms combining the excitonic and phononic contributions have also been theoretically studied~\cite{vanwezel2010,vanwezel2011,zenker2013,watanabe2015,kaneko2018}.  
Since the VB and CB are respectively composed of the Se $p$ and Ti $d$ orbitals in the different atomic sites, the band hybridization resulting from the excitonic order induces the bond formation between Ti and Se sites.  
Because electron--phonon coupling can support this spin-singlet-type Ti-Se bond formation, the excitonic bond cooperating with the lattice distortion can be expected in TiSe$_2$. 
Since the late 2000s, comparisons between experimental ARPES spectra and theoretical spectra based on the excitonic scenario have been made~\cite{cercellier2007,monney2009,monney2010,monney2010_NJP,monney2012,monney2012_NJP,monney2015}. 
It has been shown that the experimental ARPES spectra are reproducible from the theoretical spectra including electron--hole fluctuations~\cite{cercellier2007,monney2009,monney2010,monney2010_NJP,monney2012,monney2012_NJP,monney2015}. 
These comparisons between the theory and experiment became a stepping stone for the subsequent studies of TiSe$_2$ as a candidate material for the excitonic order. 

In TiSe$_2$, the phononic contribution is indispensable since the experiments have ensured the lattice distortion at low temperatures~\cite{disalvo1976,weber2011,kitou2019}.      
In other words, a pure excitonic order would not be expected in TiSe$_2$. 
Therefore, the crux of the controversy in TiSe$_2$ is how the excitonic effect contributes to the phase transition.  
The elucidation of the excitonic contribution in TiSe$_2$ is probably a topic under debate. 
In the rest of this section, we briefly introduce several works that address this issue. 
As for the study on mode softening, momentum-resolved electron energy loss spectroscopy has been applied to TiSe$_2$.
In this experiment, the softening of an electronic mode at the M (L) point associated with the phase transition is a key ingredient in the discussion of an excitonic contribution~\cite{kogar2017,lin2022}.  
Pump-probe measurements, which can visualize the real-time dynamics of nonequilibrium states, have also been applied to TiSe$_2$~\cite{rohwer2011,hellmann2012,mathias2016,monney2016,huber2022,mohrvorobeva2011,porer2014,hedayat2019,lian2020,otto2021,burian2021,duan2023}. 
Because time scales of electronic orders and lattice dynamics are usually different, pump-probe measurements can be useful tools for distinguishing between the excitonic and lattice contributions in TiSe$_2$. 
Time-resolved ARPES for TiSe$_2$ has revealed the suppression of the spectral intensity of the folded Se $p$ VB at the M (L) point and the emergence of the metallic Ti $d$ CB dispersion on a timescale as fast as 10--100 femtoseconds~\cite{rohwer2011,hellmann2012,mathias2016,monney2016,huber2022}.  
This ultrafast quenching of the CDW in a time scale of electronic dynamics is suggestive of the presence of the electronic contribution in the CDW phase. 
We expect that accurate theoretical interpretations of nonequilibrium dynamics of excitonic orders in electron--phonon coupled systems may provide conclusive insights into the driving force of the low-temperature phase of TiSe$_2$. 

Recently, similar phase transitions to TiSe$_2$ have been reported in the group-4 transition-metal ditellurides, TiTe$_2$, ZrTe$_2$, and HfTe$_2$~\cite{chen2017,gao2023,song2023,gao2024}. 
Similarly to TiSe$_2$, the ARPES measurements show the folding band at the M point reflecting the order characterized by $\bm{q}_{\rm M}$ at low temperatures.  
Interestingly, although the 2$\times$2 superlattice structures have been detected in TiTe$_2$ and ZrTe$_2$~\cite{chen2017,gao2023,song2023}, the Raman and X-ray measurements and first-principles phonon calculation for HfTe$_2$ have shown no clear signal of a lattice distortion~\cite{gao2024}.  
If the phase transition occurs without lattice distortions, its low-temperature state has the potential to be excitonic-driven.   
Elucidation of the excitonic effects in these tellurides may provide helpful insights into the phase transition in TiSe$_2$.

\subsection{Ta$_2$NiSe$_5$} \label{sec:tns}

\begin{figure}[b]
\begin{center}  
\includegraphics[width=0.95\columnwidth]{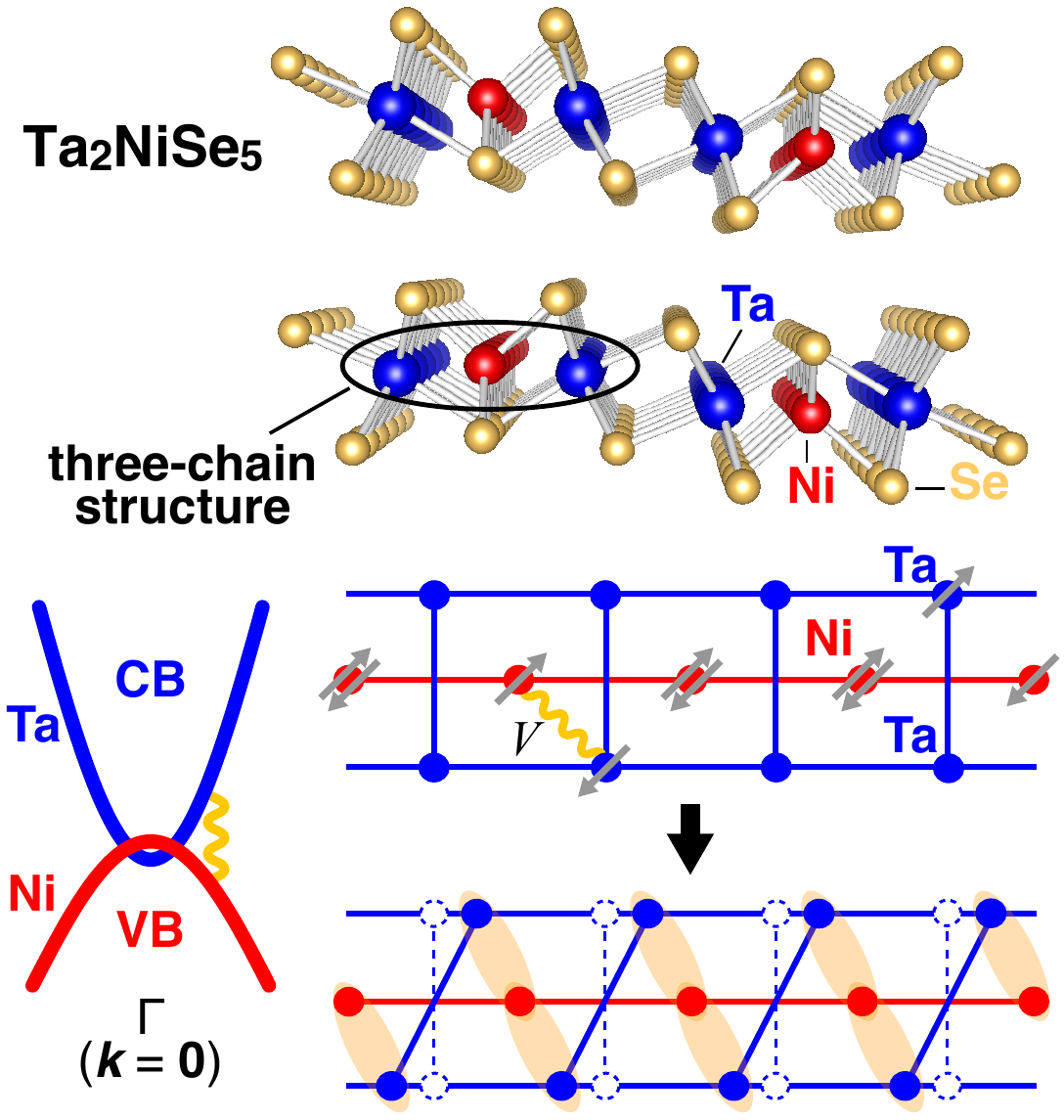}          
\caption{(Color online)
Crystal structure of Ta$_2$NiSe$_5$ visualized by VESTA~\cite{VESTA} (upper panel), schematic band structure in the normal state (lower left panel), and schematic figure of the phase transition in the three-chain structure in Ta$_2$NiSe$_5$ (lower right panel).} 
\label{fig9}
\end{center}
\end{figure}

Ta$_2$NiSe$_5$ is a material actively discussed as a candidate for the EI in the last decade. 
Ta$_2$NiSe$_5$ was composed in the 1980s~\cite{sunshine1985}. 
In Ta$_2$NiSe$_5$, six Se atoms are octahedrally coordinated around a Ta atom and four Se atoms are tetrahedrally coordinated around a Ni atom.  
Three-chain units composed of two Ta chains and one Ni chain are linked horizontally, and van der Waals layer stacking forms the bulk crystal (see the top panel of Fig.~\ref{fig9}).  
The fundamental material properties have been reported by Di Salvo {\it et al.}~\cite{disalvo1986}. 
The magnetic measurements suggest no magnetic order in Ta$_2$NiSe$_5$~\cite{disalvo1986,li2018}.  
The electric resistivity increases as temperature decreases, indicating the insulating property. 
The slope of the resistivity shows an anomaly suggesting a phase transition at around 328~K~\cite{disalvo1986,lu2017,hirose2023}. 
The X-ray diffraction studies have revealed the structural phase transition at the same temperature, where the orthorhombic structure at high temperature changes into the monoclinic structure~\cite{disalvo1986,nakano2018}.  
The bottom right panel of Fig.~\ref{fig9} schematically shows the structural transition in Ta$_2$NiSe$_5$.  
Similarly to TiSe$_2$, the phase transition in Ta$_2$NiSe$_5$ is accompanied by a lowering of the crystal symmetry. 
In contrast to the 2$\times$2 structure in TiSe$_2$, a superlattice with a renewed period is not realized in Ta$_2$NiSe$_5$. 

\begin{figure}[b]
\begin{center}  
\includegraphics[width=\columnwidth]{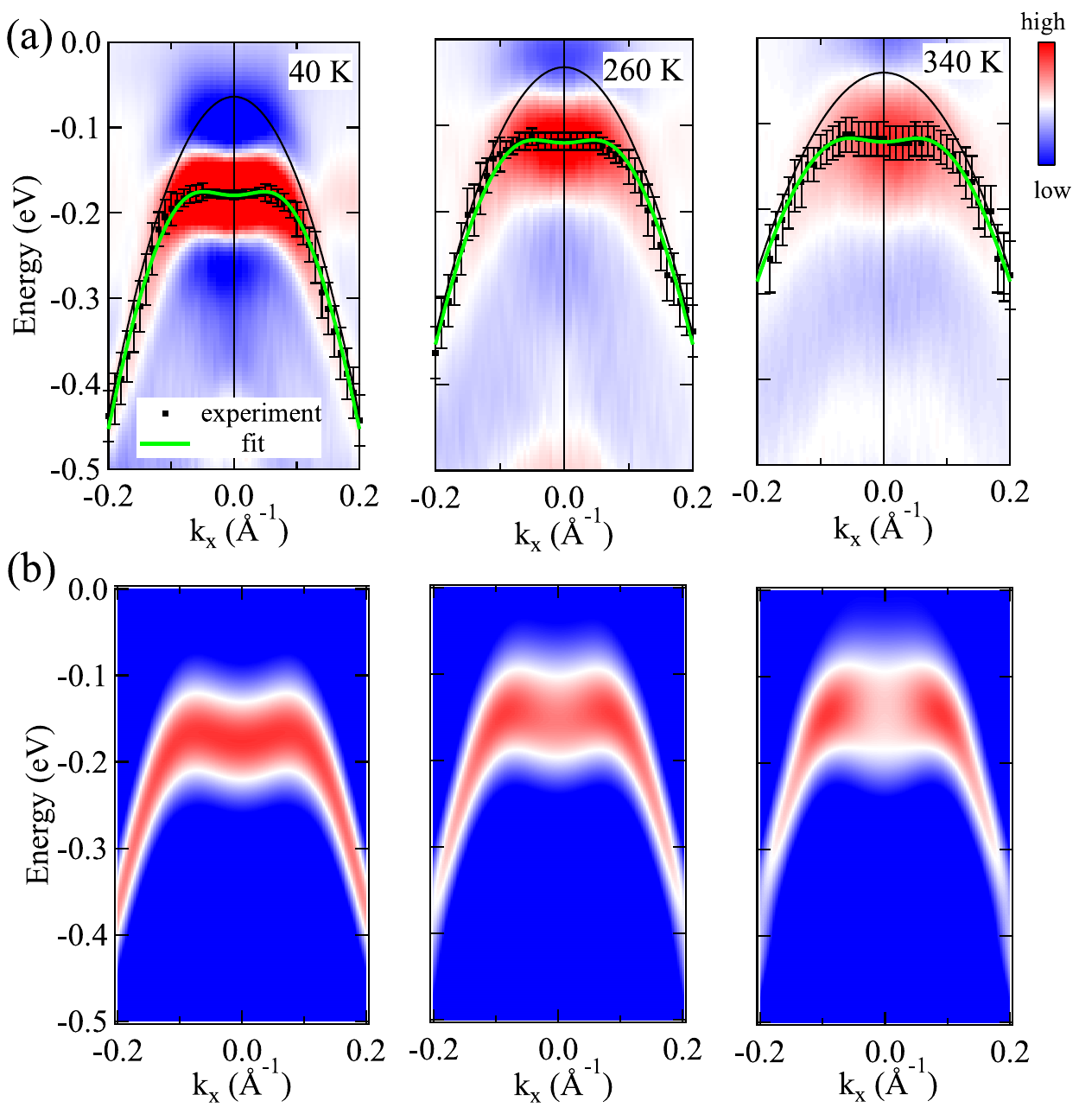}          
\caption{(Color online) 
(a) Second derivative plots of ARPES spectra of Ta$_2$NiSe$_5$ at 40, 260, and 340~K. 
(b) Single-particle excitation spectra at 40, 270, and 340~K in a quasi-1D model obtained by the VCA calculation. 
Note that contributions from electron--lattice interactions are not considered in the theoretical calculations. 
Reproduced from Ref.~\citen{seki2014} \copyright $\,$ 2014 American Physical Society.} 
\label{fig10}
\end{center}
\end{figure}

The possibility of an excitonic phase transition in Ta$_2$NiSe$_5$ was first revealed by ARPES~\cite{wakisaka2009}. 
In the ARPES experiment, the flattening of the VB top at the $\Gamma$ point, similar to hybridization gap opening in the EI, was observed in the low-temperature phase [see Fig.~\ref{fig10}(a)]. 
The flattening of the VB top at low temperatures has been reproduced by many subsequent ARPES experiments~\cite{fukutani2019,watson2020,chen2020,fukutani2021}.  
The gap opening has also been examined by scanning tunneling spectroscopy~\cite{lee2019}.  
A dome-like phase diagram corresponding to Fig.~\ref{fig2} has also been suggested as a function of pressure $P$ and doping $x$ in Ta$_2$Ni(Se$_{1-x}$S$_x$)$_5$ and Ta$_2$Ni(Se$_{1-x}$Te$_x$)$_5$~\cite{lu2017}. 
These experimental suggestions encouraged the study of Ta$_2$NiSe$_5$ as a candidate for the EI. 

The band calculations show the direct-gap-type band structure, where the VB top and CB bottom are at the $\Gamma$ point~\cite{canadell1987,mazza2020}. 
A schematic band structure is shown in Fig.~\ref{fig9} (left bottom panel). 
Because of this band structure, the flattening of the VB top observed in ARPES occurs around the $\Gamma$ point.
This characteristic should be contrasted with the ARPES spectra in the triple-$\bm{q}$ state of TiSe$_2$ because there is no folding band at other $\bm{k}$ points in Ta$_2$NiSe$_5$.  
At the $\Gamma$ point, the VB top mainly comprises the $d$ orbitals in the Ni chain while the CB bottom comprises the $d$ orbitals in the two Ta chains. 
The Ta-Ni-Ta three-chain structure shown in Fig.~\ref{fig9} is thus the main host of the physical properties~\cite{kaneko2013,mazza2020}.  
Possible excitonic scenarios for the phase transition have been theoretically proposed using simplified quasi-1D models~\cite{kaneko2013,sugimoto2016,sugimoto2016dia,yamada2016,domon2016,sugimoto2018,domon2018,yamada2019,mazza2020}. 

At the early stage, the experimental ARPES spectra are compared with theoretical spectra based on the excitonic scenario~\cite{seki2014}.  
Figure~\ref{fig10} shows the spectra obtained by the ARPES experiment and theoretical calculation~\cite{seki2014}.  
As shown in Fig.~\ref{fig10}(a), the flattening of the VB top is distinctly observed in the ARPES spectra at low temperatures, and the gap size increases as temperature decreases.  
Figure~\ref{fig10}(b) shows the single-particle excitation spectra simulated in the quasi-1D EFKM, where the VCA that can incorporate electron--hole fluctuations beyond the mean-field theory is employed. 
Note that the theoretical calculations are conducted in the purely electronic model and contributions from electron--phonon interactions are not considered. 
As seen in Fig.~\ref{fig10}, the theoretical spectra based on the EI scenario can reproduce the temperature-dependent gap opening and flattening of the VB observed in ARPES. 
In the theoretical spectra, the single-particle gap opens even without the order parameter above $T_{\rm c}$ because the electron--hole correlations that can contribute to the gap opening are incorporated in the ED-based VCA method. 
The band structures extracted from the ARPES spectra at high temperatures also do not appear to be a simple semimetallic structure~\cite{fukutani2021}, implying that the phase transition in Ta$_2$NiSe$_5$ cannot be interpreted in a simple weak-coupling BCS picture of excitonic ordering. 
Because of the flattening of the VB top at low temperatures and temperature-dependent gap opening observed in ARPES, Ta$_2$NiSe$_5$ has been discussed in the context of the EI. 

The essential issue in Ta$_2$NiSe$_5$ is similar to that of the phase transition in TiSe$_2$.
Two factors potentially contribute to the phase transition in Ta$_2$NiSe$_5$, electron--lattice coupling and excitonic interaction. 
Since the phase transition is accompanied by the lattice distortion (as schematically shown in Fig.~\ref{fig9}), electron--lattice coupling is an essential factor in symmetry breaking. 
Recent experiments have suggested the importance of the lattice effects~\cite{watson2020,baldini2023,chen2023nat,chen2023prr,haque2024}. 
Because lattice distortions can lead to the formation of atomic bonds, the lattice effect is also a key factor in describing the bonding between the Ta and Ni orbitals. 
Similarly to TiSe$_2$~\cite{vanwezel2010}, because electron--lattice coupling can support the spin-singlet-type order, the cooperative mechanism between the excitonic (interband Coulomb) and electron--lattice interactions has also been proposed in Ta$_2$NiSe$_5$~\cite{kaneko2013,sugimoto2016}.  
If the excitonic contribution is negligible in Ta$_2$NiSe$_5$ or the phase transition is purely lattice driven, we may not have to regard Ta$_2$NiSe$_5$ as an EI.  
Hence, the central controversy in Ta$_2$NiSe$_5$ is the roles of the Coulomb contribution to the phase transition.  

In contrast to TiSe$_2$, both the VB top and CB bottom are located at the $\Gamma$ point, i.e., Ta$_2$NiSe$_5$ has a direct-gap-type band structure. 
Hence, optical experiments, in which zero-momentum components of excitations can be measured, work well to capture the electronic and phononic properties in Ta$_2$NiSe$_5$. 
In the optical conductivity, the temperature-dependent gap opening associated with the phase transition has been distinctly observed~\cite{lu2017,larkin2017,okamura2023}.  
Not only the infrared-active phonons~\cite{larkin2018} but also the Raman-active phonons~\cite{kim2020,kim2021,volkov2021npj,volkov2021prb,ye2021} have been detected in the low-energy region.  
In particular, Raman spectroscopy has shown the asymmetric lineshape of the phonon mode. 
This Raman spectrum is interpreted as the excitation continuum, which prominently emerges near the transition temperature, overlaps with the phonon modes~\cite{kim2021,volkov2021npj}.   
The peak position of the phonon mode related to the structural transition does not simply shift to low energies, whereas the spectral weight of the broad continuum near zero energy is maximized at the transition temperature~\cite{kim2021,volkov2021npj}.  
This anomalous softening behavior in Ta$_2$NiSe$_5$ suggests that the phase transition is not simply phonon-driven but involves a fluctuation of an electronic entity~\cite{kim2021,volkov2021npj}.  

Since optical experiments can capture the key material properties of Ta$_2$NiSe$_5$, light-induced nonequilibrium dynamics have also been measured in Ta$_2$NiSe$_5$.  
For example, time-resolved ARPES has visualized the fast deformation of the electronic band structures on the order of 100~fs~\cite{mor2017,okazaki2018,mitsuoka2020,tang2020,suzuki2021,saha2021,golez2022,takahashi2023} and the oscillation of the spectral weight reflecting the excitation of the coherent phonons on the order of 1--10~ps~\cite{tang2020,suzuki2021}.    
The contribution of the excitonic interaction appearing in time-resolved ARPES has been discussed, such as in Refs.~\citen{saha2021} and \citen{golez2022}, by comparing the experimental nonequilibrium spectra with theoretical spectra obtained in effective models. 
The optical pump-probe measurements have also shown the ultrafast deformation of the transient optical response function and the oscillation of spectra associated with the coherent phonon excitation~\cite{werdehausen2018,werdehausen2018JPh,mor2018,ning2020,bretscher2021nat,bretscher2021sci,miyamoto2022,guan2023,katsumi2023,takamura2024,jiang2024}. 
The interplay between the lattice effects and possible excitonic contribution has been actively discussed to interpret the measured optically driven dynamics in Ta$_2$NiSe$_5$. 
Although not everything would have been fully elucidated to date because complex lattice and electronic band structures of Ta$_2$NiSe$_5$ cause theoretical difficulties in obtaining the exact solution in nonequilibrium states, the experimental findings are steadily accumulating. 
As we discuss later (see Sect.~\ref{sec:CM}), we expect that the elucidation of the nonequilibrium dynamics of the EI candidates will provide important insights. 

Interestingly, Ta$_2$NiSe$_5$ shows superconductivity under pressure~\cite{matsubayashi2021}.  
The emergence of superconductivity is common to TiSe$_2$, which shows superconductivity under pressure~\cite{kusmartseva2009}, Cu intercalation~\cite{morosan2006}, and controlling carrier density~\cite{li2016}.  
In Ta$_2$NiSe$_5$, pressure suppresses the phase transition accompanied by structural distortion.  
Although the pressure-driven structural change at $P\sim 3$~GPa slides a layer in a unit cell, Ta$_2$NiSe$_5$ in the high-pressure phase also shows the orthorhombic-monoclinic structural transition at low temperatures~\cite{nakano2018IUCrJ}. 
The superconducting dome emerges when the transition temperature to the low-temperature monoclinic phase approaches zero. 
Similar temperature-pressure phase diagrams are plotted in TiSe$_2$~\cite{kusmartseva2009}. 
The phase diagrams of TiSe$_2$ and Ta$_2$NiSe$_5$ recall the phase diagrams of cuprate~\cite{keimer2015} and iron-based superconductors~\cite{hosono2015}, whereas the parent electronic orders are crucially different. 
The origin of superconductivity in these EI candidates is an interesting issue. 

Recently, the possibility of an EI state has been suggested in Ta$_2$Pd$_3$Te$_5$~\cite{huang2024,zhang2024,hossain_arxiv}, which has a similar layered structure to Ta$_2$NiSe$_5$. 
The ARPES experimental results for Ta$_2$Pd$_3$Te$_5$ show a semimetallic band structure at high temperatures but indicate a gap opening at low temperatures. 
In contrast to Ta$_2$NiSe$_5$, the structural distortion in Ta$_2$Pd$_3$Te$_5$ is tiny, and it is expected that the lattice effect is not a main contributor to the gap opening~\cite{huang2024,zhang2024,hossain_arxiv}.   
We expect further experimental and theoretical studies on this material in the future.

\subsection{Co oxides} \label{sec:co_based}

As a candidate for a spin-triplet-type excitonic order, we introduce several cobalt oxides with the $d^6$ configuration. 
Cobalt oxides with the perovskite-type crystal structure have long been investigated as a system exhibiting the spin-state transition (or crossover) between the spin states of Co ions.  
For example, the electronic state of LaCoO$_3$ can be divided into three regions depending on temperature: a diamagnetic insulator at low temperatures ($T\lesssim$ 80~K), a paramagnetic insulator at intermediate temperatures, and a paramagnetic metal at high temperatures ($T\gtrsim$ 600~K), where no clear phase transitions have been observed between these regions, but they change in a crossover manner~\cite{heikes1964,raccah1967,korotin1996,mizokawa1996,haverkort2006,eder2010,tomiyasu2017,takegami2023}.  
The competition between the low-spin (LS, $S = 0$), intermediate-spin (IS, $S = 1$), and high-spin (HS, $S = 2$) states, which occurs due to the delicate balance between the crystal field splitting and ferromagnetic Hund's rule coupling of Co$^{3+}$ (3$d^6$), has been considered to govern the physical properties of this system~\cite{heikes1964,raccah1967,korotin1996,mizokawa1996,haverkort2006,eder2010,tomiyasu2017,takegami2023}. 
However, despite more than half a century of research, the entire picture still needs to be clarified. 
Then, a possibility has been raised that an excitonic phase will appear in the vicinity of these states. 
In the following, taking Pr$_{0.5}$Ca$_{0.5}$CoO$_3$ and LaCoO$_3$ as examples, we will give an overview of the possibility of realizing an excitonic phase in which a spin-triplet excitonic order is formed between different orbitals within one atom. 
We will, in particular, discuss Pr$_{0.5}$Ca$_{0.5}$CoO$_3$ without a magnetic field and LaCoO$_3$ under a high magnetic field.  

First, let us consider Pr$_{0.5}$Ca$_{0.5}$CoO$_3$. 
This material exhibits a first-order metal-insulator transition at $T\sim90$~K with a decrease in magnetic susceptibility and a sharp peak in specific heat~\cite{tsubouchi2002,tsubouchi2004,fujita2004,hajtmanek2010,hardy2013,hajtmanek2013}. 
In the low-temperature phase, no magnetic moment is observed so that no trivial magnetic long-range order can be identified, resulting in what we call a hidden order. 
It is known that the valence transition from Pr$^{3+}$ to Pr$^{4+}$ occurs simultaneously~\cite{hajtmanek2013}. 
The valence of Co is close to Co$^{3+}$ ($3d^6$) at low temperatures~\cite{hajtmanek2013}. 
Despite the lack of magnetic order, exchange splitting is observed in the Kramers doublet of the $4f$ orbital of Pr$^{4+}$~\cite{hajtmanek2010,hajtmanek2013}. 

\begin{figure}[t]
\begin{center}  
\includegraphics[width=0.5\columnwidth]{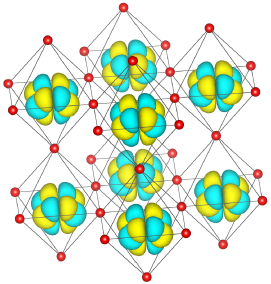}          
\caption{(Color online) 
Magnetic multipole structure predicted in Pr$_{0.5}$Ca$_{0.5}$CoO$_3$ as a consequence of a spin-triplet-type excitonic order. 
Reproduced from Ref.~\citen{yamaguchi2017} \copyright $\,$ 2017 The Physical Society of Japan.} 
\label{fig11}
\end{center}
\end{figure}

Then, from the analyses of the TOHM using the DMFT, together with the results of the band-structure calculations using LDA+$U$, Kune\v{s} and Augustinsk\'{y}~\cite{kunes2014Co} suggested that the phase transition in Pr$_{0.5}$Ca$_{0.5}$CoO$_3$ should be due to a spin-triplet excitonic order, where the magnetic multipole structure as a consequence of the excitonic ordering was proposed for Pr$_{0.5}$Ca$_{0.5}$CoO$_3$. 
Also, Yamaguchi {\it et al}.~\cite{yamaguchi2017,yamaguchi2018} applied the mean-field approximation to the three-dimensional five-orbital model obtained for Pr$_{0.5}$Ca$_{0.5}$CoO$_3$ and demonstrated the formation of the magnetic multipole order associated with the triplet excitonic order (see Fig.~\ref{fig11}). 
A similar estimation was also conducted by using the LDA+$U$ calculation in the $d^6$ cobalt oxide~\cite{afonso2017}.  
Because net magnetization in a unit is nearly zero in the magnetic multipole structure shown in Fig.~\ref{fig11}, the excitonic order proposed in Pr$_{0.5}$Ca$_{0.5}$CoO$_3$ can behave as a hidden order. 
In addition, Yamaguchi {\it et al}.~\cite{yamaguchi2017,yamaguchi2018} calculated the generalized magnetic susceptibility using the random phase approximation for the five-orbit model and found the characteristic low-energy collective-mode excitations in the spin-triplet excitonic phase. 
Their possible observability in inelastic neutron scattering experiments has also been discussed;~\cite{yamaguchi2017,yamaguchi2018} actually, an attempt was made to verify the presence of this mode~\cite{moyoshi2018}. 
As for a spectral study, the resonant inelastic X-ray scattering (RIXS) experiment has also been conducted to determine the excitonic dispersion of the IS spin state of LaCoO$_3$~\cite{wang2018}. Further experimental verification of the excitonic order in Co-oxide systems predicted in the theoretical studies is desirable.  

Next, let us consider LaCoO$_3$ under a high magnetic field~\cite{ikeda2024}. 
The high magnetic field is a valuable tool for studying field-induced changes in the spin states of materials. 
Theoretical studies based on multiorbital models have predicted a variety of phases including excitonic ordered phases under a magnetic field~\cite{sotnikov2016,sotnikov2017,tatsuno2016}. 
For LaCoO$_3$, a strong pulsed magnetic field above 100~T was applied to this system, and magnetization and magnetostriction measurements were performed over a wide temperature range~\cite{ikeda2016,ikeda2020}.  
A similar experiment was conducted for Y-doped Pr$_{0.7}$Ca$_{0.3}$CoO$_3$~\cite{ikeda2016Pr}.  
More recently, ultrahigh magnetic fields up to 600~T have been applied, and magnetostriction measurements were performed~\cite{ikeda2023}.  
Thus, the phase diagram in the temperature and magnetic-field plane was proposed~\cite{ikeda2023}, which includes a variety of phases with different spatial spin orders. 

\begin{figure}[t]
\begin{center}  
\includegraphics[width=0.95\columnwidth]{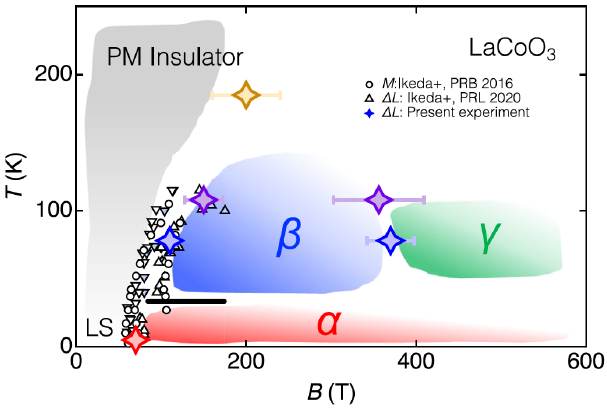}          
\caption{(Color online) 
Temperature--magnetic-field phase diagram of LaCoO$_3$ obtained from the magnetostriction measurement under ultrahigh magnetic fields. 
Reproduced from Ref.~\citen{ikeda2023} \copyright $\,$  2023 The Authors, CC BY 4.0 license.} 
\label{fig12}
\end{center}
\end{figure}

Figure~\ref{fig12} shows the temperature--magnetic-field phase diagram of LaCoO$_3$ presented by Ikeda {\it et al.}~\cite{ikeda2023}. 
Under a low magnetic field, a diamagnetic phase at low temperatures and a paramagnetic phase at intermediate temperatures exist in a crossover manner, which connects to the phases of LaCoO$_3$ at zero fields. 
As the strong magnetic field is applied, after passing through a broad first-order phase-transition region, two field-induced phases appear: the $\alpha$ phase at low temperatures ($<20$~K) in the entire range of the magnetic field applied and the $\beta$ phase at intermediate temperatures (between $\sim$20 and $\sim$120~K) above $\sim$380~T~\cite{ikeda2023}. 
Then, by further increasing the magnetic field beyond $\sim$380~T, a new phase denoted as $\gamma$ appears at around $\sim$80~K~\cite{ikeda2023}. 
The possible origins of the $\alpha$, $\beta$, and $\gamma$ phases were proposed to be as follows:~\cite{ikeda2023} the $\alpha$ phase is a superlattice of spatially localized bi-excitons and vacuum, the $\beta$ phase is a phase of spatially uniform excitonic order, and the $\gamma$ phase is described as a superposition of the superlattice of bi-excitons and the spatially uniform excitonic order. 
These results are based on some theoretical calculations where the mean-field analysis was performed on an effective model derived from the five-orbital Hubbard model on the cubic lattice~\cite{ikeda2023}.  
As introduced in this section, Co oxides are attractive platforms for studying excitonic phases, and further investigations are desirable to clarify their detailed properties.

\subsection{Other candidate materials} \label{other_candidates}

Other intriguing proposals on excitonic phases have so far been made for various materials. 
Although not all materials can be introduced in detail, we will list as many materials as possible. 
We note that controversial materials and theoretical proposals that have not been experimentally established may be included in the following. 

EI states have been studied in atomic layer materials. 
An important example is monolayer WTe$_2$~\cite{wang2021,jia2022,sun2022}, which has been discussed to be a quantum spin Hall insulator as one of the topological insulators, where the reported quantum oscillation of the resistance has been explained in terms of gap modulation in the hybridized Landau levels of an EI~\cite{kwan2021,lee2021,he2021}.  
A signature of the gap opening at low temperatures has also been reported in this system~\cite{jia2022,sun2022,que2024}.  
As theoretically predicted~\cite{varsano2020}, monolayer TMDs with the $T'$ structure have the potential to be a topological EI~\cite{varsano2020}, and the interplay between topology and excitonic correlations is an interesting aspect of these materials.  

Another well-known example may be TmSe$_{1-x}$Te$_x$, where a pressure--temperature phase diagram similar to the dome-like structure shown in Fig.~\ref{fig2} has been proposed~\cite{neuenschwander1990,bucher1991,wachter2004}. 
The spontaneous hybridization between the conduction 5$d$ band and the valence 4$f$ band of Tm$^{2+}$ ions has been suggested to occur, explaining the possible presence of an excitonic phase under high pressures~\cite{neuenschwander1990,bucher1991,wachter2004}.  
As for a material involving $f$ electrons, GdGaI has recently been proposed as an EI~\cite{okuma_arxiv}.  
GdGaI exhibits a similar band structure to that of TiSe$_2$, and a replica band due to band folding at low temperatures has been revealed by the ARPES experiment~\cite{okuma_arxiv}.  
It has been expected that the electronic bands near the Fermi level composed of the Gd 5$d$ orbitals are coupled to the localized Gd 4$f$ spins, and the gap opening interacts with the triple-$\bm{q}$ magnetic order of the localized $f$ spins~\cite{okuma_arxiv}.    

As predicted by Volkov {\it et al.}~\cite{volkov1975}, ferromagnetism can occur when extra electrons are doped into EI states. 
Measurements on electron-doped calcium hexaboride (CaB$_6$) showed that its ground state is ferromagnetically polarized with a saturation moment of $0.07\mu_B$ and an ordering temperature of 600~K~\cite{young1999}.  
Soon after this finding, theoretical calculations~\cite{zhitomirsky1999,balents2000prl} were reported on the doped CaB$_6$, which showed that ferromagnetism with a small magnetic moment but with a high Curie temperature can be obtained by doping a spin-triplet EI state. 
However, we note that it was pointed out after careful experiments that the reported ferromagnetism was not of excitonic origin but instead of extrinsic origin, which can be ascribed to iron impurity phases~\cite{matsubayashi2002,matsubayashi2003}. 

Carbon-based materials, such as carbon nanotubes~\cite{varsano2017}, graphite under high magnetic field~\cite{zhu2019}, and semihydrogenated graphene,~\cite{jiang2020} have been discussed as candidates for EIs. 
In particular, the last graphene material has been predicted to be a spin-triplet EI~\cite{jiang2020}.  
A diatomic kagome lattice with a flat band predicted as a spin-triplet excitonic state~\cite{serthi2021,sethi2023,delgado_arxiv} may also be among this category. 
Various theoretical predictions have recently been made for the possible presence of EI states. 
In MoS$_2$ under high pressure, the first-principles many-body perturbation theory has predicted the presence of an ideal EI phase~\cite{samanehataei2021}.  
Excitonic effects have been discussed in a nodal line semimetal ZrSiS, where the unusual mass enhancement of charge carriers observed experimentally~\cite{pezzini2018} has been explained as the excitonic instability and pseudogap formation~\cite{rudenko2018, scherer2018}.  
Another material is 1$T$-$MX$$_2$ ($M$=Co, Ni and $X$=Cl, Br), where the unusual electronic state dubbed as half-EI has been predicted to occur, in which one spin channel has a many-body ground state due to excitonic instability and the other is characterized by a conventional band-insulator gap~\cite{jiang2019,disabatino2023}.  
GaAs with the double-layer honeycomb structure has been proposed as a direct-gap-type EI~\cite{jiang2018}. 
A transition-metal trichalcogenide TiS$_3$ has also been discussed as a candidate, where a possible high transition temperature is predicted in a monolayer~\cite{dong2020,wang2021TiS3}.  
Recently, the possibilities of excitonic orders have been proposed in kagome metals~\cite{scammell2023}.
The topological EI states have recently been discussed in various 2D systems~\cite{wang2019,liu2021,sdong2023,yang2024}.  
Possible excitonic states have also been suggested in the TMD moir\'e bilayer systems~\cite{xie2023,zdong2023,xie2024}. 
We hope that further experimental studies will be made on these materials to demonstrate the presence of such EI states.  

Excitonic physics has also been studied in transition-metal oxides containing $4d$ and $5d$ electrons, in which phenomena are related to spin-orbit coupling. 
Khaliullin~\cite{khaliullin2013} proposed the concept of excitonic magnetism, which is caused by the condensation of excitons formed between holes with effective angular momentum $J_{\rm eff} = 3/2$ and electrons with $J_{\rm eff} = 1/2$ in a $d^4$ electron system placed in a cubic crystal field. 
The spin-singlet ground state of an atom itself does not lead to particularly interesting low-energy physics. 
However, if the spin-triplet excitation carrying the magnetic moment is the lowest-energy excitation and has sufficiently low energy, the exchange interaction between atoms caused by electron hopping leads to an ordered state. 
Owing to the exchange process between singlets and triplets, the triplet excitation can be considered a quasiparticle moving in a singlet background with nontrivial dispersion. 
If the lowest-energy level of this dispersion comes into contact with the singlet level, the system undergoes a phase transition. 
The spin-orbit coupling must be sufficiently large to allow this excitation, with the gap between singlet and triplet being small enough. 
Although we will not go into detail in this review, the feasibility of this excitonic magnetism in real materials has been discussed for Ir$^{5+}$-based double perovskite materials such as Sr$_2$YIrO$_6$, Ba$_2$YIrO$_6$, and A$_2$BIrO$_6$ (A=Ba, Sr; B=Lu, Sc)~\cite{cao2014,dey2016,pajskr2016,kusch2018,lagunamarco2020,aczel2022}, for which future developments will be interesting. 
In addition, the feasibility of excitonic magnetism due to the ordering of similar triplet excitons has been discussed in the Ru oxide Ca$_2$RuO$_4$ with a layered perovskite structure~\cite{akbari2014,jain2017,yamamoto2022}, providing a broad perspective on the excitonic physics in correlated electron systems. 

Another important example recently reported may be Sr$_3$Ir$_2$O$_7$, which has been argued to be a realization of the antiferromagnetic spin-triplet EI~\cite{mazzone2022,suwa2021}.  
A bilayer single-orbital Hubbard model, which is an effective model for the system with spin-orbit coupling and exchange interactions, has been used to explain the observed electronic and magnetic properties of the material~\cite{mazzone2022,suwa2021}.  
We may also add a $(t_{2g})^5$ electron system, where the EI state may occur by combining spin-orbit interactions through a mechanism different from that described above~\cite{sato2015}.  

Possible EI phases have also been considered in multilayer devices, where the carrier densities of each layer can be controlled exogenously. 
In this respect, semiconductor bilayer systems such as InAs/GaSb~\cite{du2017,yu2018} under the bias voltage are discussed as a topological EI~\cite{du2017}.  
Also, the EI states involving interlayer excitons have been examined in the bilayer systems of TMDs such as MoSe$_2$/WSe$_2$~\cite{ma2021,nguyen_arxiv}.  
The electron--hole systems combining  moir\'e TMD bilayers have also been discussed as hosts of EI states~\cite{chen2022,zhang2022,gu2022,xie2023,zeng2023}.  
Recently, the possibility of dynamical exciton condensation has been investigated in electron--hole bilayer systems~\cite{zeng2024,sun2024}. 
If we broaden the topic to exciton condensation in light-matter coupled systems, we can find various references, such as exciton-polariton condensation in semiconductor microcavities~\cite{kasprzak2006,deng2010}.  
We will not discuss such states further because they are beyond the scope of this article that mainly focuses on the EIs appearing as the unbiased ground state of materials.


\section{Collective Modes in Excitonic Insulators} \label{sec:CM}

In the candidate materials TiSe$_2$ and Ta$_2$NiSe$_5$, the phase transitions are accompanied by lattice distortions. 
Since the lattice contributions hinder the identification of the excitonic contribution in these materials, it is often debated whether the deformations of the electronic states are due to the lattice distortions or excitonic ordering.
If contributions from electron correlations to a low-temperature phase are substantial, a material can be regarded as an EI. 
In other words, if excitonic contributions are negligible and a lattice distortion is the main cause of a low-temperature phase, there is no need to regard such material as an EI. 
When we consider an excitonic order and electron-lattice coupling at the level of the static single-particle approximation, their parameters are simultaneously incorporated in the off-diagonal elements of the multiband Hamiltonian, and distinguishing these two contributions becomes difficult in the simple band structure.  
This is because not only excitonic ordering but also band hybridization due to lattice deformations can generate the expectation value $\braket{\hat{a}^{\dag}\hat{b}}$. 
When two contributions work cooperatively, the issue of which contribution comes first often becomes a chicken-and-egg problem.  
The issue in the candidates TiSe$_2$ and Ta$_2$NiSe$_5$ is similar to the controversy over whether the metal-insulator transition in vanadium oxides is a Mott transition due to electronic correlations or a Peierls transition due to lattice distortions~\cite{grandi2020}. 

The dynamics of the excitonic entity and lattice structure, i.e., their behavior in excited states, may be important in separating these two factors. 
Since the electronic and lattice systems usually have different time scales in excitation phenomena, elucidating their contributions in excited states can provide a clue for solving controversial problems. 
In the case of excitations of lattice vibrations, we investigate the phonon properties. 
On the other hand, if we consider excitations originating from an excitonic order, we must consider collective excitations caused by fluctuations of the order parameter~\cite{murakami2017,murakami2020,remez2020,lenk2020,sun2020,golez2020,sun2021topo,sun20212nd}. 
Here, we review the collective properties of the EI coupled to lattice degrees of freedom. 
First, we introduce the excitation profile of the pure excitonic order. 
Then, we consider the effects of electron--lattice coupling on the collective modes of the excitonic order. 
Finally, we comment on their optical properties to provide perspectives on observability.

\subsection{Collective modes of excitonic orders} \label{sec:CM_EI}

\begin{figure}[b]
\begin{center}  
\includegraphics[width=\columnwidth]{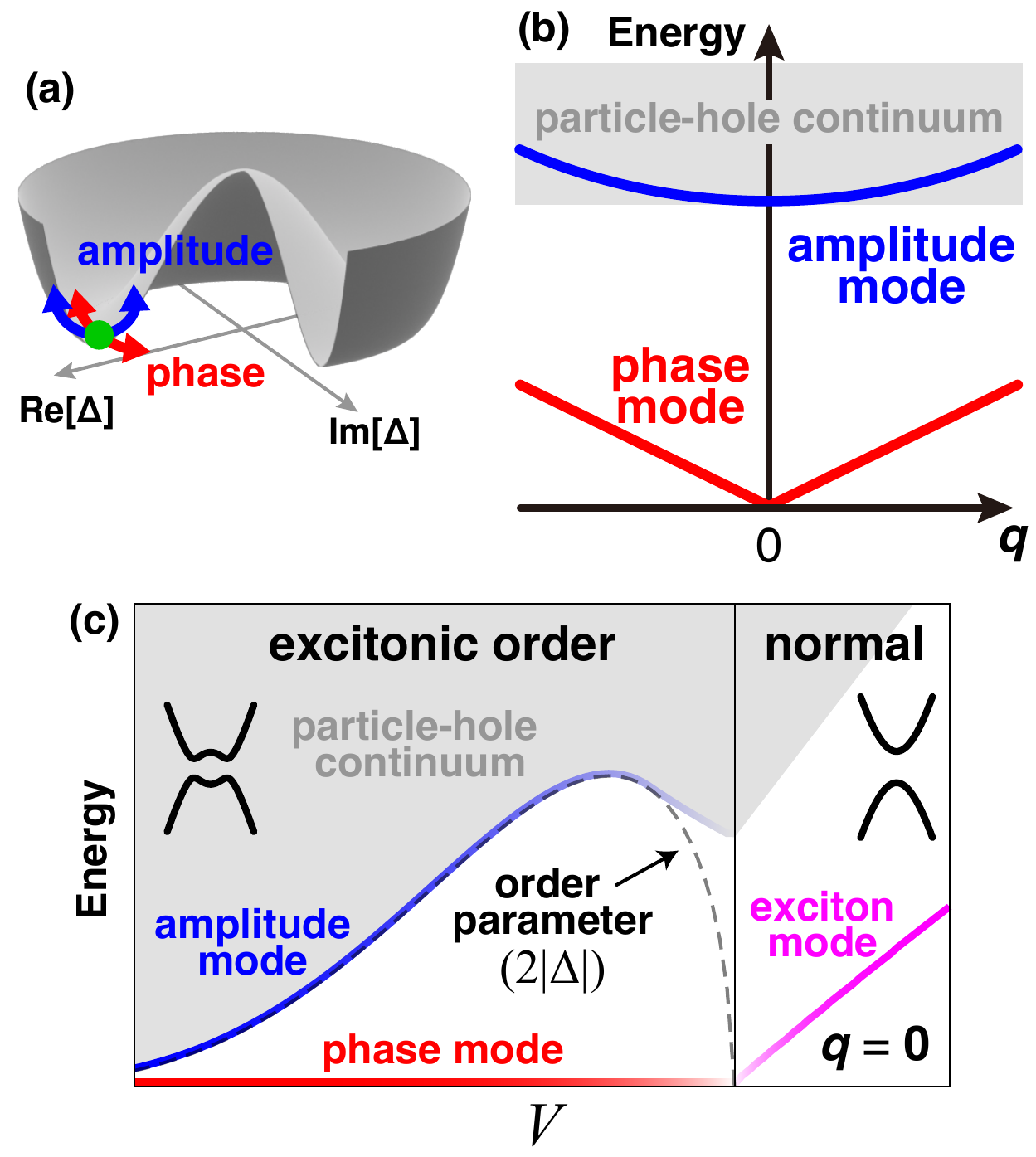}          
\caption{(Color online) 
(a) Schematic figure of the amplitude and phase fluctuations on the energy landscape of the pure excitonic order.
(b) Momentum ($\bm{q}$)-dependent collective excitation modes in the pure excitonic order, where the shadowed region indicates the particle-hole continuum above the single-particle gap ($\sim  2|\Delta|$ in the weak-coupling BCS regime).  
(c) Parameter ($V$) dependence of the excitation spectrum at $\bm{q}=\bm{0}$. 
The dashed line in the ordered phase corresponds to the magnitude of the order parameter.} 
\label{fig13}
\end{center}
\end{figure}

First, we review the collective excitations of the pure excitonic order. 
For simplicity, we consider the spinless two-band correlated model composed of the Hamiltonians of Eqs.~(\ref{eq:ham_twoband_0}) and (\ref{eq:ham_twoband_V}) with the direct-gap-type band structure. 
(The collective modes in the models with spin degrees of freedom have been studied in, for example, Refs.~\citen{nasu2016} and \citen{geffroy2019}.)
The collective excitations in the excitonic order originate from the fluctuations of the order parameter $\Delta$. 
In general, the order parameter $\Delta$ can be a complex number because it is included in the off-diagonal elements of the matrix Eq.~(\ref{eq:hammf_twoband_matrix}).  
The complex order parameter can be described as $\Delta = |\Delta| e^{i\theta}$, where the order parameter $\Delta$ is decomposed into the amplitude $|\Delta|$ and the phase $\theta$. 
In the symmetry-breaking ground state, the amplitude is nonzero, i.e., $|\Delta|>0$, and the phase $\theta$ is fixed at one point.  
Since the difference of $\theta$ does not change the eigenenergy of the pure excitonic order shown in Eq.~(\ref{eq:EI_eigenenergy}), the fixed point of the phase is arbitrary.  
This corresponds to the U(1) symmetry breaking.  
The deviation of the order parameter from its equilibrium value gives rise to the collective excitation of the excitonic order.   
Specifically, $|\Delta| \rightarrow |\Delta| + \delta \Delta$ gives the amplitude mode while $\theta \rightarrow \theta + \delta \theta$ gives the phase mode. 
As in another ordered state with spontaneous symmetry breaking, the amplitude mode corresponds to the Higgs mode while the phase mode corresponds to the Nambu--Goldstone mode. 

Figure~\ref{fig13} shows a typical excitation structure in the pure excitonic order.  
In addition to the collective excitation, there are single-particle (independent particle) excitations across the gap. 
In the weak-coupling BCS regime, the gap is given by $2|\Delta|$. 
In Figs.~\ref{fig13}(b) and \ref{fig13}(c), the excitation continuum attributed to the single-particle excitations is represented by the shadowed region. 
As shown in Fig.~\ref{fig13}(a), the energy landscape of the ordered state is often compared to the Mexican hat shape. 
In the pure excitonic order, the excitation along the phase direction does not take energy because the energy landscape is isotropic [see Fig.~\ref{fig13}(a)].  
The phase mode is thus gapless, as shown in Fig.~\ref{fig13}(b). 
On the other hand, the excitation due to the change of the amplitude $|\Delta|$ requires energy. 
In the pure excitonic order, the bottom of the amplitude mode usually coincides with the bottom of the excitation continuum, as shown in Fig.~\ref{fig13}(b).  

\begin{figure}[b]
\begin{center}  
\includegraphics[width=0.7\columnwidth]{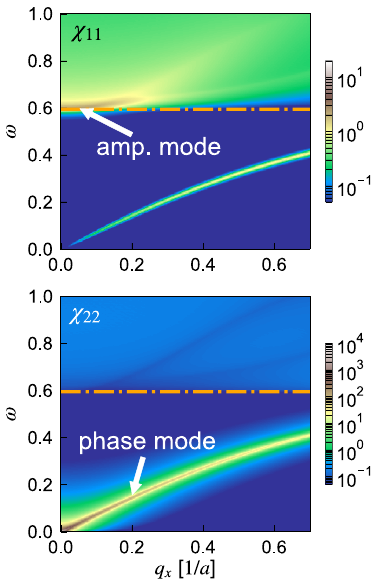}          
\caption{(Color online) 
Linear response functions of the order parameter $\bm{\chi}_{11}(\omega;\bm{q})$ and $\bm{\chi}_{22}(\omega;\bm{q})$, where $- (1/\pi) {\rm Im} \chi_{\mu\mu}(\omega;q_x,q_y=0)$ in an anisotropic 2D model is plotted. 
The dashed-dotted lines indicate the energy of the single-particle gap.  
The hopping integral along the $x$ direction is used as the unit of energy. 
Reproduced from Ref.~\citen{murakami2020} \copyright $\,$ 2020 American Physical Society.} 
\label{fig14}
\end{center}
\end{figure}

We can find this excitation structure in the dynamical correlation functions of the excitonic order parameter~\cite{murakami2020}.  
To describe the excitation spectra, we introduce the operator $\hat{\rho}^{\mu}_j= \sum_{\gamma,\gamma'} \hat{\gamma}^{\dag}_{j} \tau^{\mu}_{\gamma\gamma'} \hat{\gamma}'_{j}$, where $\gamma,\gamma' = a, b$ and $\tau^{\mu}$ ($\mu=0,1,2,3$) is the Pauli matrix.   
$\hat{\rho}^{1}_j=\hat{a}^{\dag}_j \hat{b}_j + \hat{b}^{\dag}_j \hat{a}_j$ represents the real part of the order parameter. 
The $\mu=2$ component represents the imaginary part of the order parameter while the $\mu=3$ component represents the difference in the particle filling. 
The dynamical linear response function for these operators is given by $\chi_{\mu\nu}(t-t';\bm{r}_{ij})=-i\theta(t-t') \braket{[\hat{\rho}^{\mu}_{i}(t), \hat{\rho}^{\nu}_{j}(t')]}$, where $\theta(t)$ is the step function and $\bm{r}_{ij}=\bm{r}_i - \bm{r}_j$. 
The excitation spectra are obtained by the Fourier transform of $\chi_{\mu\nu}(t-t';\bm{r}_{ij})$. 
Within the time-dependent mean-field approximation (see Ref.~\citen{murakami2020} for details), the linear response function in the frequency and momentum space is given by 
\begin{align}
\chi(\omega;\bm{q}) = \chi^{(0)}(\omega;\bm{q}) + \chi^{(0)}(\omega;\bm{q}) \Theta(\omega;\bm{q}) \chi(\omega;\bm{q}), 
\end{align}
where $\chi(\omega;\bm{q})$ and $\chi^{(0)}(\omega;\bm{q})$ are 4$\times$4 matrices for the $\mu,\nu=0,1,2,3$ components and $ \Theta(\omega;\bm{q})={\rm Diag}[V/2,-V/2,-V/2,-V/2]$. 
The bare response function $\chi^{(0)}(\omega;\bm{q}) $ is given by 
\begin{align}
\chi^{(0)}_{\mu\nu} (\omega;\bm{q}) 
= \frac{1}{N} \sum_{\bm{k}} & \sum_{\alpha,\,\beta=\pm}  {\rm Tr} \left[ W_{\alpha}(\bm{k}-\bm{q}) \tau^{\mu} W_{\beta}(\bm{k}) \tau^{\nu} \right]
\notag \\
& \times \frac{f(E_{\alpha}(\bm{k}-\bm{q})) - f(E_{\beta}(\bm{k}))}{\hbar \omega + i \eta - \left[ E_{\beta}(\bm{k}) - E_{\alpha}(\bm{k}-\bm{q}) \right]}, 
\end{align}
where $W_{\pm}(\bm{k})=(1/2) [ \tau^0 \pm \bm{\tau} \cdot \bm{B(\bm{k})}/| \bm{B}(\bm{k}) | ]$ with $B^1(\bm{k})=-2\rm{Re} \Delta$, $B^2(\bm{k})=-2\rm{Im} \Delta$, and $B^3(\bm{k})=\bar{\varepsilon}_a(\bm{k}) - \bar{\varepsilon}_b(\bm{k})$~\cite{murakami2020}. 
$E_{\alpha}(\bm{k})$ corresponds to the energy in Eq.~(\ref{eq:EI_eigenenergy}) obtained by the diagonalization of matrix (\ref{eq:hammf_twoband_matrix}). 
$\hbar$ is the Planck constant and $\eta$ is a small positive number.  
Figure~\ref{fig14} shows the linear response function $-(1/\pi) {\rm Im} \chi_{\mu\mu}(\omega;\bm{q})$ in the BCS regime when the order parameter is set to real ($\theta=0$) in equilibrium~\cite{murakami2020}. 
In this case, $\chi_{11} (\omega;\bm{q})$ mainly reflects the amplitude fluctuation while $\chi_{22} (\omega;\bm{q})$ mainly reflects the phase fluctuation~\cite{murakami2020}.   
As seen in Fig.~\ref{fig14}, the excitation continuum exists above the single-particle gap $2|\Delta|$, and the gapless mode grows from $\omega=0$ and $\bm{q}=\bm{0}$. 
$\chi_{22} (\omega;\bm{q})$ shows a strong spectral weight around $\omega=0$ and $\bm{q}=\bm{0}$, indicating that this gapless mode corresponds to the phase mode. 
On the other hand, the spectral weight of $\chi_{11} (\omega;\bm{q})$ is strongly enhanced at the bottom of the excitation continuum ($\hbar \omega \sim 2 |\Delta|$). 
This enhanced spectrum around $\hbar\omega=2|\Delta|$ and $\bm{q}=\bm{0}$ corresponds to the presence of the amplitude mode, as shown in Fig.~\ref{fig13}(b).   
The signature of the amplitude mode is distinct in the BCS regime but becomes less prominent in the BEC regime~\cite{murakami2020}.  

To understand the relation between the phase transition and the excitation spectrum, we schematically show the parameter ($V$) dependence of the collective mode at $\bm{q} = \bm{0}$ in Fig.~\ref{fig13}(c). 
In the model of Eq.~(\ref{eq:ham_twoband_V}), the gap between the VB top and the CB bottom increases with $V$ because of the energy level shift due to the Hartree term of the interaction. 
Hence, in the large $V$ region, the ground state is a normal semiconductor (band insulator) with a gap due to the Hartree shift [see the inset of Fig.~\ref{fig13}(c)]. 
Although there is no excitonic order in this normal semiconductor phase, there is a spectral peak due to the electron--hole bound state, i.e., exciton, formed by the Coulomb interaction $V$. 
As indicated by the pink line in Fig.~\ref{fig13}(c), when the band gap is sufficiently large, the exciton mode appears at the energy $E_{\rm g} - |E_{\rm B}|$ that is lower than the gap energy $E_{\rm g}$ between the VB top and the CB bottom (i.e., the bottom of the continuum), where $E_{\rm B}$ corresponds to the exciton binding energy. 
However, when the band gap is small and becomes equal to the exciton binding energy (i.e., $E_{\rm g}=|E_{\rm B}|$), the energy level of the exciton peak reaches zero. 
Then, the phase transition to the excitonic order occurs [see Fig.~\ref{fig13}(c)]. 
This corresponds to the softening of the exciton mode. 
The point of softening coincides with the point at which the order parameter becomes nonzero. 
In the excitonic phase, there is a phase mode (red line) at zero energy corresponding to the $\bm{q}=\bm{0}$ point in Fig.~\ref{fig13}(b). 
Additionally, at the bottom of the excitation continuum, i.e., the energy gap due to the order parameter ($2|\Delta|$), there is a spectral enhancement, as shown in Fig.~\ref{fig14},  corresponding to the presence of the amplitude mode.
Therefore, the phase and amplitude modes of the order parameter exist in the excitonic phase, which is triggered by the softening of the exciton mode in the semiconductor phase.

\subsection{Effects of electron--phonon coupling}

The excitation properties discussed in the previous section are valid only for the pure excitonic order. 
In actual materials, other factors such as electron--lattice coupling potentially contribute to band hybridization (off-diagonal element in the multiband matrix) aside from the excitonic order parameter. 

\begin{figure}[b]
\begin{center}  
\includegraphics[width=0.9\columnwidth]{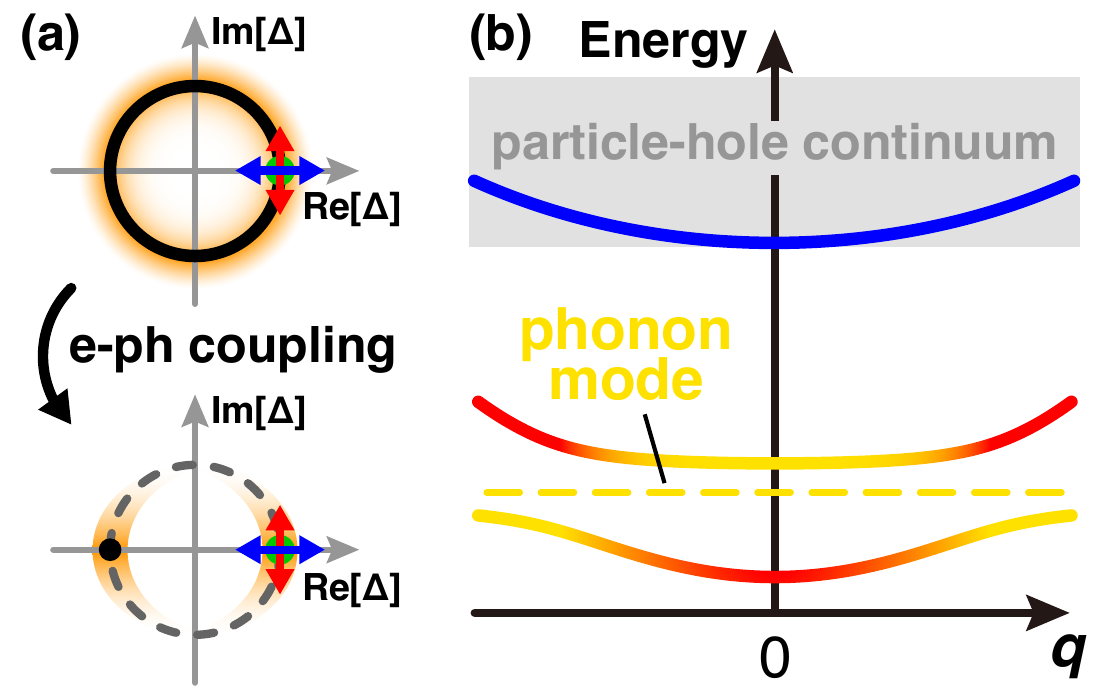}          
\caption{(Color online) 
(a) Schematic figure of phase locking caused by electron--phonon coupling. 
The thick black line in the upper panel indicates the lowest energy in the energy landscape of the pure excitonic order. 
The black dots in the lower panel indicate the lowest energy points in the electron--phonon coupled system.  
The horizontal and vertical arrows represent the amplitude and phase fluctuations, respectively.  
(b) Collective excitation modes in the excitonic order with electron--phonon coupling, where the dashed line indicates a bare phonon mode that couples to the phase mode of the excitonic order.} 
\label{fig15}
\end{center}
\end{figure}

Here, we consider the effects of the electron--lattice coupling.   
In this section, we assume the simple electron--lattice coupling $g \sum_{j} x_j ( \hat{a}^{\dag}_j \hat{b}_j + \hat{b}^{\dag}_j \hat{a}_j)$ employed in Ref.~\citen{murakami2020}, where $x_j$ corresponds to a displacement at site $j$. 
Note that actual forms of electron--lattice coupling are more complicated and depend on materials. 
The excitonic order parameter and lattice distortion pattern are usually uniform in the direct-gap-type band structure shown in Fig.~\ref{fig1}. 
Considering this electron--lattice coupling, the off-diagonal element of the Hamiltonian becomes $\Delta + gx$, indicating that the lattice contribution changes the energy landscape.  
When $gx$ is real, the phase of the order parameter $\Delta$ is fixed at $\theta=0$ (or $\theta=\pi$ with $gx<0$) to minimize the energy~\cite{murakami2020}.  
Figure~\ref{fig15}(a) schematically shows the phase locking in the electron--phonon coupled system. 
This results from the symmetry reduction from U(1) to Z$_2$. 
Hence, the phononic contribution can crucially change the properties of the collective excitations. 

The behavior of the collective excitations with the effect of a phonon is schematically shown in Fig.~\ref{fig15}(b). 
When the single-particle gap is larger than the phonon energies, the phonon modes exist at an energy below the single-particle excitation continuum. 
The excitation structure shown in Fig.~\ref{fig15}(b) assumes a single phonon mode that can couple to the excitonic order parameter.  
The dispersive phase mode of the excitonic order also exists below the single-particle gap. 
Then, the phase and phonon modes are hybridized via electron--phonon coupling. 
As a result, hybridized collective modes are created, as shown in Fig.~\ref{fig15}(b). 
In addition to the formation of the hybrid modes, the lowest excitation mode becomes gapped [see Fig.~\ref{fig15}(b)].  
This is because the phase $\theta$ of the order parameter is fixed by electron--lattice coupling [see Fig.~\ref{fig15}(a)].   
Since the energy is minimized at the fixed point involving electron--lattice coupling, its excitation requires energy, resulting in a gap in the collective mode.   
Meanwhile, although the bottom of the amplitude mode usually coincides with the bottom of the excitation continuum, it is known that orbital hybridization attributed to the hopping and nonlocal Coulomb interactions in real materials can shift the peak position of the amplitude mode to an energy different from the bottom of the continuum~\cite{murakami2020,kaneko2021}. 

\begin{figure}[t]
\begin{center}  
\includegraphics[width=0.9\columnwidth]{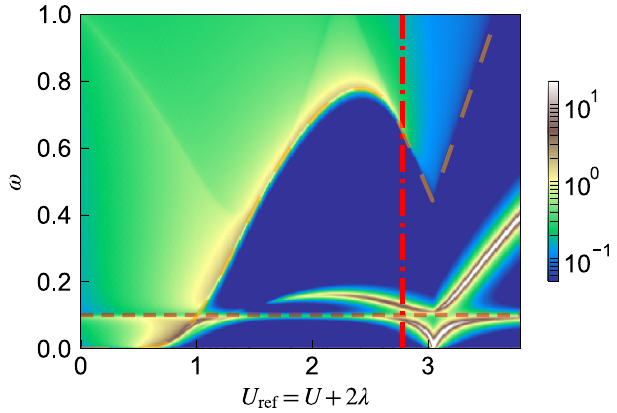}          
\caption{(Color online) 
Linear response function $-(1/\pi) {\rm Im} \chi_{11}(\omega;\bm{q}=\bm{0})$ in the EI with electron--phonon coupling, where $U$ is equal to $V$ and $\lambda$ $\propto g^2/\omega_0$ corresponds to the strength of electron--phonon coupling.  
The horizontal red dashed line represents the bare phonon frequency, and the dashed orange line indicates the single-particle gap. 
The vertical dotted-dashed red line indicates a guideline for semiconductor-semimetal crossover (see Ref.~\citen{murakami2020} for details). 
Reproduced from Ref.~\citen{murakami2020} \copyright $\,$ 2020 American Physical Society.} 
\label{fig16}
\end{center}
\end{figure}

Figure~\ref{fig16} shows the parameter dependence of $\chi(\omega;\bm{q}=\bm{0})$, taking into account the electron--lattice coupling. 
In the dynamical linear response function $\chi(\omega;\bm{q})$, the phonon contribution $g^2 D^{(0)}(\omega)$, where $D^{(0)}(\omega)= 2 \hbar \omega_0/ [ (\hbar \omega + i \eta)^2 - \hbar^2 \omega_0^2]$ is the free-phonon Green function with the phonon frequency $\omega_0$, is added to the 11 component of $\Theta(\omega;\bm{q})$ (see Ref.~\citen{murakami2020} for details). 
Even in the normal semiconductor phase (without the order parameter), the exciton mode is hybridized with the phonon mode, and these hybridized exciton-phonon modes exist in the sub-bandgap excitation regime.  
As shown in Fig.~\ref{fig16}, the softening of one of the hybrid modes induces the phase transition and causes the ordered state. 
In the ordered phase, an optimal ground state is formed by coupling between the excitonic order and the lattice system. 
The hybridized entities of the phonon and collective excitonic phase modes exist in the sub-bandgap excitation regime (see Fig.~\ref{fig16}).  
The peak positions of the lower and upper hybrid modes in the excitation spectrum depend on the strength of the electron--lattice coupling~\cite{murakami2020}.  
A strong electron--lattice coupling tends to push the upper hybrid mode to higher energy. 
If we can detect these collective modes in experiments, we may extract the strength of the phonon contribution. 

As briefly shown here, even if it is difficult to find a difference between the excitonic order and lattice contributions in the static band structure, there are differences in features between the collective modes of the excitonic order and the phonon excitation. 
The modes possessing different characteristics form the hybridized modes via the electron--lattice interactions. 
Therefore, it is important to investigate the excitation dynamics of the candidate materials, such as Ta$_2$NiSe$_5$, on the basis of a good understanding of the collective excitonic properties considering the electron--lattice couplings. 
These perspectives will help elucidate the origin of the phase transitions involving the lattice deformations. 

Although we considered the collective modes in the simple spinless model, electron spin is also an intrinsic factor in real materials.
For example, the pair-hopping interaction in the multiorbital Hubbard model results in a gapful phase mode~\cite{nasu2016}. 
Because additional interactions in spinful models can modify the energy landscape, the assessment of collective modes involving the full electron--electron interactions and electron--phonon couplings is an important issue for a more accurate discussion in real materials.

\subsection{Optical responses} \label{sec:EI_CM_OR}

To investigate responses attributed to collective excitations, experiments with the application of some external field to a target system are required.  
As introduced in Sect.~\ref{sec:candidates}, the responses to light have been studied extensively in the candidate materials TiSe$_2$ and Ta$_2$NiSe$_5$. 
In recent years, pump-probe spectroscopy has been used to measure the real-time dynamics of nonequilibrium states, and nonlinear optical responses driven by strong external fields have also enabled us to obtain valuable insights. 
The dynamics of the order parameters under the light field are important in understanding the optical responses of the EI candidates. 
Since light-induced phenomena depend on the optical selection rule between the VB and CB, the light-induced dynamics of the order parameter strongly depend on the electronic features near the Fermi level.  

We can readily see the collective excitations of the excitonic modes when the order parameter is directly stimulated by light.  
For example, when the valence and conduction orbitals have different parities and the dipole $\bm{d}$ characterized by the overlap of two atomic orbitals is nonzero [e.g., $d_x \propto \int d\bm{r} \psi_s(\bm{r}) x \psi_{p_x}(\bm{r}) \ne 0$ between the $s$ and $p_x$ orbitals], the optical excitation can be induced by the term $\bm{E}(t)\cdot \bm{d} \, \hat{a}^{\dag}_j \hat{b}_j + {\rm H.c.}$, where $\bm{E}(t)$ is the electric field.  
In this case, the electric field is introduced in the off-diagonal elements including the order parameter, i.e., light can directly stimulate the excitonic order parameter~\cite{murakami2020,murakami2017}.  
In this case, because the optical conductivity is proportional to the linear response function $\chi_{11}(\omega;\bm{q}=\bm{0})$ (see Ref.~\citen{murakami2020} for details), we can directly detect the collective modes originating from the excitonic order and phonon in the linear response regime. 
Hence, if the off-diagonal excitation is directly allowed, we may easily find the collective phenomena in optical experiments.  

However, the presence of the dipole $\bm{d}$ depends on the material. 
When $\bm{d}=\bm{0}$, we have to extract collective properties in optical responses induced by the electric field introduced via the Peierls substitution $\bm{k} \rightarrow \bm{k} -  (q/\hbar) \bm{A}(t)$, where $q$ is the charge of a particle and $\bm{A}(t)$ is the vector potential.
When the external field is only introduced via this procedure in the model of Eqs.~(\ref{eq:ham_twoband_0}) and (\ref{eq:ham_twoband_V}) with the spatial inversion symmetry, the collective modes of the excitonic order do not appear in the linear optical responses (at $\bm{q}=\bm{0}$)~\cite{murakami2020}.  
This is because the order parameter is not activated by light in the linear response regime (when $\bm{d}=\bm{0}$), i.e., the linear optical response function is not corrected by the dynamical order parameter~\cite{murakami2020,tanabe2021}.  
Since second-order nonlinear optical responses are prohibited in inversion-symmetric systems, third-order optical spectroscopy becomes a reporter for the order parameter dynamics. 
Theoretically, it has been proposed that the collective dynamics of the excitonic order can be captured by third-harmonic generation (THG), in which the $3\Omega$ response is driven by incident light with the frequency $\Omega$~\cite{tanabe2021}. 
In particular, the THG response function (in the BCS regime) is enhanced by the amplitude mode~\cite{tanabe2021}, indicating the possibility of the direct measurement of changes in the amplitude of the order parameter using light.
This property has some analogy to the amplitude (Higgs) mode of superconductors~\cite{shimano2020}. 
In superconductors, enhanced THG responses have been observed by terahertz spectroscopy~\cite{matsunaga2014,shimano2020}. 
When the order parameter of the excitonic order is much larger than the energy scale of phonon modes, we can expect the separation of the amplitude-mode dynamics and phonon effects.
This suggests that measurements of nonlinear optical responses to the amplitude mode can be useful for revealing the origin of the low-temperature phase of the candidate materials.  
On the other hand, when the inversion symmetry is broken in the EI ground state, the linear response function can capture the collective mode of the excitonic order~\cite{kaneko2021}. 
This is because the order parameter can be activated by light in the linear response regime due to the absence of the inversion symmetry. 

The pumped nonequilibrium dynamics of EIs under various driving procedures have been theoretically investigated~\cite{golez2016,murakami2017,tanaka2018,tanabe2018,khatibi2020,tuovinen2020,fujiuchi2019,ejima2022,perfetto2019,perfetto2020prl,perfetto2020prb}.  
The gap closure and light-induced enhancement of the order parameter have been demonstrated by the time-dependent mean-field (Hartree--Fock) approximation and the ansatz beyond it~\cite{golez2016,murakami2017,tanaka2018,tanabe2018,khatibi2020,tuovinen2020}.    
The melting of the excitonic correlation and the time profile of the spectral features in the optically driven EFKM have been studied by the ED-based 
and matrix-product-state (MPS)-based numerical time-evolution techniques~\cite{fujiuchi2019,ejima2022}. 
Nonequilibrium EIs have been proposed in pumped band insulators~\cite{perfetto2019,perfetto2020prl,perfetto2020prb}.   
In terms of collective modes in an EI coupled with lattice degrees of freedom, Gole\v{z} {\it et al.}~\cite{golez2020} theoretically demonstrated that nonlinear excitation resonant to the phase mode induces a new in-gap mode at twice the phase-mode frequency when the electronic contribution is dominant for the formation of the ground-state order.  
The detection of characteristics different from those in the lattice-dominant case in observable nonlinear spectroscopy may be a smoking gun of an excitonic order. 

In real materials, as introduced in Sect.~\ref{sec:candidates}, various pump-probe spectroscopies are applied to the ordered phases in the candidate materials, TiSe$_2$ and Ta$_2$NiSe$_5$. 
In particular, since the direct-gap-type material Ta$_2$NiSe$_5$ is suitable for optical experiments, many collaborations between experimental and theoretical groups have targeted this material to elucidate its nonequilibrium dynamics. 
For example, the deformations of the band structure observed in time-resolved ARPES have been compared with the time profile of the pumped EI described in simple theoretical models considering electron--phonon coupling to examine the excitonic contribution in Ta$_2$NiSe$_5$~\cite{mor2017,saha2021,golez2022}.  
The micrometer-scale propagation of excited modes detected in pump-probe microscopy has been interpreted as the propagation of the collective excitation of the hybridization mode between a phonon and the phase mode of the excitonic order parameter~\cite{bretscher2021sci}.  
Light-induced transition in Ta$_2$NiSe$_5$ has been discussed in association with the ultrafast reversal of an Ising order parameter ($\Delta \rightarrow -\Delta$) in an energy landscape with electron--lattice coupling~\cite{ning2020,guan2023}.   
Although simplified electron--phonon coupled models have been used for comparison with experimental results, theoretical studies of pumped band structure and collective modes using realistic electronic bands and a lattice distortion pattern are gradually developing~\cite{geng2024,chatterjee_arxiv}. 
We expect that cooperation between experiment and theory will be further strengthened, and the full picture of the collective dynamics of EI materials will be revealed in the future.


\section{Summary and Outlook} \label{sec:summary}

We reviewed the recent progress of the studies on the EI. 
First, we introduced the concept of the EI using a simple theoretical model and discussed its relation to the BCS--BEC crossover. 
We also reviewed the theoretical development of the research of the EIs in the context of strongly correlated electron systems, i.e., the EI states studied in the EFKM and TOHM using various numerical techniques.   
Then, we introduced the candidate materials for the EI that have attracted attention in the last few decades. 
In particular, we focused on the ordered states in TiSe$_2$ and Ta$_2$NiSe$_5$, where the phase transitions are accompanied by lattice distortions, as candidates for the spin-singlet-type excitonic order. 
We also introduced several cobalt oxides near the spin-state crossover regime as candidate hosts for the spin-triplet-type excitonic order. 
Finally, we discussed the collective modes of the excitonic order coupled to the lattice system.  
We suggested the importance of the collective dynamics of the EI to elucidate the excitonic contribution in phase transitions involving lattice distortions. 

The EI system contains many essences of condensed matter physics. 
For example, order formation is often discussed in terms of the physics of condensation of fermion pairs in a BCS--BEC crossover manner. 
An EI involving charge, spin, and orbital degrees of freedom can be a platform for emerging rich physical properties. 
Depending on the orbital textures of the VB and CB, an excitonic order can create an electrical/magnetic multipole structure or bond formation across atomic sites. 
Although theory has been advanced for a long time, the rise of new candidate materials and the development of experimental techniques have enabled us to discuss EI states in cooperation between theoretical and experimental groups. 
This is an important advancement in the last few decades. 

However, debatable issues would remain in identifying evidence of the EI state. 
Besides the excitonic order parameter, other contributions to the gap opening exist in real materials.  
For example, in TiSe$_2$ and Ta$_2$NiSe$_5$, the phase transitions involve lattice distortions. 
The lattice contribution hinders the identification of the excitonic contributions because the excitonic interaction and electron--phonon coupling can cause the ordered states.  
As discussed in Sect.~\ref{sec:CM}, we expect that studies of the collective properties in these systems play a key role in detecting the excitonic effects. 
Pump-probe measurements gradually reveal the dynamical properties of the candidate materials. 
From this perspective, an electron--hole interacting system without lattice distortions may be a promising candidate for the EI. 

Since the Coulomb interaction is the driving force of the EI, strongly correlated materials configured with an even number of electrons (e.g., $d^{2n}$) may be suitable systems for the EI. 
When the VB and CB are composed of orbitals in the same atom, local on-site Coulomb interactions can be a driving force of excitonic order. 
The cobalt oxides with the $d^6$ configuration introduced in Sect.~\ref{sec:co_based} are systems suited to this phenomenon. 
We expect that new candidate materials hosted on strongly correlated systems will be discovered in the future. 
Flat band systems such as those realized by moir\'{e} engineering in twisted bilayers, can also have a large effective interaction against the bandwidth. 
Systems with tunable band gaps, interactions, and particle filling will also be suitable platforms for studying the EI.

Although not all topics have been covered in detail, we hope that this review article will help enhance the reader's understanding.


\section*{Acknowledgments}
We thank S.~Ejima, H.~Fehske, R.~Fujiuchi, D.~Gole\v{z}, K.~Hamada, M.~Kadosawa, T.~Konishi, A.~J.~Millis, S.~Miyakoshi, Y.~Murakami, H.~Nishida, S.~Nishimoto, K.~Seki, K.~Sugimoto, Z.~Sun, T.~Tanabe, T.~Toriyama, M.~Udono, T.~Yamaguchi, S.~Yamamoto, S.~Yunoki, and B.~Zenker for fruitful collaborations. 
We also thank H.~Fukuyama, S.~Ishihara, M.~Itoh, N.~Katayama, Y.~Kobayashi, K.~Matsubayashi, T.~Mizokawa, J.~Nasu, A.~Nakano, Y.~Nemoto, Y.~Okada, H.~Okamura, K.~Okazaki, Y.~\={O}no, M.~Sato, M.~Suzuki-Sakamaki, H.~Sawa, H.~Takagi, and Y.~Wakisaka for constructive discussions. 
This work was supported by Grants-in-Aid for Scientific Research from JSPS, KAKENHI Grant Nos. JP20H01849, JP24K06939, and JP24H00191.


\bibliography{references}

\end{document}